\documentclass[10pt,journal,compsoc]{IEEEtran}
\usepackage[nocompress]{cite}
\usepackage[dvips]{graphicx}
\graphicspath{{./eps/}}
\DeclareGraphicsExtensions{.eps}
\usepackage{amsmath}
\usepackage{amssymb}
\usepackage[ruled]{algorithm2e}
\usepackage{array}
\usepackage[caption=false,font=footnotesize,labelfont=sf,textfont=sf]{subfig}
\usepackage{stfloats}
\usepackage{url}
\usepackage{balance}
\usepackage{hyphenat}
\usepackage{multirow}

\begin{document}
%
%

\title
	{
	A Design Space Exploration Methodology for Parameter Optimization in Multicore Processors
	}
%
%
\author
  {
  Prasanna~Kansakar,%
  ~\IEEEmembership{Student~Member,~IEEE}
  and~Arslan~Munir,%
  ~\IEEEmembership{Member,~IEEE}

  \IEEEcompsocitemizethanks
    {
    \IEEEcompsocthanksitem The authors are with the Department of Computer Science, Kansas State University, Manhattan, KS\protect\\
e-mail: \{pkansakar@ksu.edu,~amunir@ksu.edu\}
    }
	}
\sloppy
\nohyphens{	
%
%
\IEEEtitleabstractindextext
  {
  \begin{abstract}
  The need for application-specific design of multicore/manycore processing platforms is evident with computing systems finding use in diverse application domains. In order to tailor multicore/manycore processors for application specific requirements, a multitude of processor design parameters have to be tuned accordingly which involves rigorous and extensive design space exploration over large search spaces. In this paper, we propose an efficient methodology for design space exploration. We evaluate our methodology over two search spaces – small and large, using a cycle-accurate simulator (ESESC) and a standard set of PARSEC and SPLASH-2 benchmarks. For the smaller design space, we compare results obtained from our design space exploration methodology with results obtained from fully exhaustive search. The results show that solution quality obtained from our methodology are within 1.35\% - 3.69\% of the results obtained from fully exhaustive search while only exploring 2.74\% - 3\% of the design space. For larger design space, we compare solution quality of different results obtained by varying the number of tunable processor design parameters included in the exhaustive search phase of our methodology. The results show that including more number of tunable parameters in the exhaustive search phase of our methodology greatly improves solution quality.
  \end{abstract}
%
%
  \begin{IEEEkeywords}
  multicore/manycore processors, processor design parameters, design space exploration, cycle-accurate simulator (ESESC), PARSEC and SPLASH-2 benchmarks, parameter optimization
  \end{IEEEkeywords}
  }
\maketitle
%
%
\IEEEraisesectionheading{\section{Introduction and Motivation}
\label{introduction_and_motivation}}
\IEEEPARstart{C}{omputing} technology is used in several diverse application domains each having different application-specific requirements. These requirements can seldom be efficiently met by generically designed computing systems. So, for different application domains, application-specific systems have to be designed such that domain-specific system requirements are met, while staying within the boundaries of feasible design. The design of application-specific multicore/manycore processors involves tuning of settings for processor design parameters to find a nearly-optimal design configuration which best meets application-specific requirements. There are two main challenges that need to be addressed in the process of tuning of settings for processor design parameters -- efficiently exploring the design space, and, satisfying multiple (possibly conflicting) design metrics.

An efficient design space exploration methodology is a must because of constraints of time and resources. Even with the processing and memory capabilities of current high-end machines, simulating all the possible configurations in a large design space is not temporally feasible. Design time is an important factor that needs to be considered in development of processors because the longer the design time of the processor, the longer the time-to-market for it will be. The longer time-to-market can result in significant revenue loss for the vendor if the product does not enter the market window (i.e., period during which the product would have highest sales) on-time \cite{fvESD01}.

The design space exploration methodology must also be able to find solutions when multiple conflicting design metrics \cite{jbMOO08} need to be satisfied. When considering optimization of multiple conflicting metrics, it is not possible to have global optimal solution (i.e., design configuration) in which all of the conflicting metrics have optimal values. Instead, the best solution is the a trade-off that can be obtained between the given conflicting metrics. The design problem then can be modelled as an Optimal Production Frontier problem also known as Pareto Efficiency \cite{sbCO04} problem. In such a problem, several trade-off solutions are obtained where each solution favors one of the conflicting metrics. The choice of solution from the set of trade-offs can be made based on the application-specific requirement of the design.

In this paper, we propose a methodology for multicore/manycore processor parameter optimization, which partially explores the design space to determine a parameter configuration which gives the best trade-off between the specified application requirements. The approach used in this paper, utilizes a combination of exhaustive, greedy and one-shot (initial tunable parameter settings selection) searches to efficiently perform design space exploration. Our methodology can efficiently prune the design space, and can accommodate multiple conflicting design metrics in the optimization process. We extensively tested our methodology on a cycle-accurate simulator, ESESC (Enhanced Super ESCalar), using a large set of multi-threaded benchmarks, PARSEC (Princeton Application Repository for Shared-Memory Computers) and SPLASH-2 (Stanford ParalleL Applications for SHared memory, version 2).

The main contributions of our paper are:
\begin{itemize}
  \item We propose a methodology for multicore parameter tuning that combines three different exploration methods, exhaustive searching, greedy searching and one-shot searching, to prune the design space to find the best settings for tunable design parameters to meet particular application requirements.
  \item We present an initial optimization algorithm based on one-shot searching that can determine initial settings for each tunable parameter to within 51.26\% of the best setting for that parameter.
  \item We describe a set partitioning algorithm, which uses results from the initial one-shot optimization algorithm, to group parameters based on the significance they have towards the targeted application-specific design requirements.  We use three significance groups in our methodology one for each of the exploration methods, exhaustive, greedy and one-shot searching.
  \item Our set partitioning algorithm includes an exhaustive search threshold factor, which allows the designer to control design time by manipulating the number of design parameters considered in the exhaustive search.
  \item We propose exhaustive and greedy search algorithms, which improve on the initial settings obtained by the initial optimization process, to yield best settings to within 1.35\% - 3.69\% of the best settings obtained from fully exhaustive search of the design space.
  \item We compare the effects of varying the exhaustive search threshold factor on the quality of solutions obtained from our methodology.
\end{itemize}
The remainder of the paper is organized as follows. Section \ref{related_work} gives a review of related work. Section \ref{methodology} describes our methodology for parameter optimization. Section \ref{algorithm} presents the algorithms leveraged by our parameter optimization methodology. The experimental setup describing the simulator and benchmarks used to test the algorithm is presented in Section \ref{experimental_setup}. Section \ref{results} discusses the results. Finally, Section \ref{conclusion_and_future_work} concludes our study along with a brief description of future research directions.
%
%
\section{Related Work}
\label{related_work}
There has been work done in literature relevant to processor parameter optimization \cite{rkHMCA03, mmPPTDSEMA08, xqCCP12} and several innovative optimization methodologies have been proposed. Several research articles are available in which authors have used design space exploration algorithms such as exhaustive search, greedy search, genetic algorithms, evolutionary algorithms etc. We present a brief analysis of the outcomes obtained by some authors from these different approaches.

A fully exhaustive search of the design space is the ideal method of design space exploration as it will certainly lead to the best design configuration, but, the overhead involved in performing such a search limits its usability. Much research has been carried out to devise methods of narrowing the scope of exhaustive search to form an equally effective partial exhaustive search algorithm. One of such methods was proposed by Givargis et~al. \cite{tgPLATUNE02}, in their system PLATUNE (PLATform TUNEr), used to simulate parameterized SoCs (System on Chip) for embedded applications. Their algorithm was separated into two phases. In the first phase, they searched through the design space to find strongly interdependent parameters and grouped them into clusters. Exhaustive searches were carried out on each cluster separately to determine the pareto-optimal configuration for each cluster. These configurations were termed locally pareto-optimal configurations. In the second phase, to extend the search over the complete design space to find globally pareto-optimal configurations, exhaustive search was carried out only on the locally pareto-optimal configurations obtained from each cluster. Their system could prune a design space as large as $10^{14}$ configurations, but took on order of 1-3 days for completing the search.

Palesi et~al. \cite{mpDSEGA02}, improved on the work presented by Givargis et~al. \cite{tgPLATUNE02}. They argued that the system PLATUNE was feasible only when the number of strongly interdependent parameters in each cluster was small. This is true because if too many parameters are clustered together, then, the partial search-space of each cluster will be large enough to make exhaustive search infeasible. To overcome this, Palesi et~al. introduced a new threshold value in their exploration algorithm which distinguished between clusters based on the size of their partial search-space. If the size of the partial search-space for a cluster was smaller than the threshold value, then, exhaustive search was used. However, if the size of the partial design space was greater than the threshold value, then, instead of using an exhaustive search over the cluster, a genetic exploration algorithm was used. The same distinction was extended for searching through the entire design space. Through this improvement, they were able to achieve 80\% reduction in simulation time while still remaining within 1\% of the results obtained from exhaustive search.

Genetic algorithms were also used in the system MULTICUBE (Multi-objective Design Space Exploration of MultiProcessor-SoC Architectures for Embedded Multimedia Applications), by Silvano et~al. \cite{csMC10}. In this system, they defined an automatic design space exploration algorithm that could swiftly present an approximate pareto-front to the system designer to help in deciding which design configuration was best suited for a particular application scope. The exploration algorithms used in their work range from different variants of genetic algorithms to evolutionary algorithm and simulated annealing. In their work, they presented a comparison of these different exploration algorithms based on how well they converge towards the optimal design configuration based on the percentage of design space explored. They argued that the degree of closeness to the best setting for each tunable parameter in a design space is strictly related to the number of evaluations that the system designer can afford to make. In other words, the degree of closeness to the best configuration is directly proportional to the percentage of design space explored.

Munir et~al. \cite{amWSN13} used a greedy algorithm to overcome the overhead of exhaustive search, in their paper on dynamic optimization of wireless sensor networks. Their algorithm was separated in two phases. In the first phase, a one-shot search algorithm was employed to find the best initial values for each of the tunable parameters considered. The search was limited to the first and last settings within the set of possible settings for each parameter. Once the best initial values for all the parameters were determined, the parameters were ordered based on the significance that each parameter has towards the targeted application-specific requirement. In the second phase, a greedy search algorithm was used which worked off of the best initial values determined in the first phase. The greedy search, progressed through the list of possible settings for each parameter, starting at the best initial value. If the new setting for a parameter, yielded a better configuration than the previous setting for that parameter, then, the search was continued. But, if the new setting, yielded a configuration that is worse than the previous configuration, then, the search algorithm was terminated for that parameter and the same search algorithm was started for the next parameter. In their paper, they compared the performance of their algorithm against the performance of simulated annealing exploration algorithm. They concluded that their algorithm converged to within 8\% of the best configuration while only exploring 1\% of the design space as compared the simulated annealing exploration algorithm that explored 55\% of the design space to get within the same range of convergence.

Some of the other widely used approaches to prune design space include the use of machine learning algorithms and statistical simulation. Guo et~al. \cite{tcDSEUDC14} used machine learning in their design space exploration algorithm. In their system, a training set was formed using a small number of design configurations. The training set was simulated and their simulation results were used to generate a predictive model. Once an accurate predictive model was obtained it was used to predict the simulation results of design configurations not in the training set. This greatly reduced exploration time as simulations were only carried out on the small training set. Genbrugge et~al. \cite{dgDSESS09} used statistical simulation in their design space exploration algorithm. They reduced exploration time by reducing the size of their simulations. They achieved this by generating synthetic trace of a benchmark program's execution by using statistical profiling. The synthetic trace had the same execution characteristics as the benchmark program but with far smaller simulation time.

In this paper, we improve on the work carried out by Munir et~al. \cite{amWSN13}. We use a similar approach to design space exploration but with an addition of two new phases - a set partition phase and an exhaustive search phase. With the addition of the exhaustive search phase we intend to increase the degree of closeness to the optimal solution by exploring a larger portion of the design space, as argued by Silvano et~al. \cite{csMC10}. The limit on the number of configurations considered in the exhaustive search is determined by the set-partitioning phase in which we use the threshold concept presented by Palesi et~al. \cite{mpDSEGA02}.
%
%
\section{Parameter Optimization Methodology}
\label{methodology}
%
%
\subsection{Overview}
\label{methodology_overview}
\begin{figure*}[!t]
  \centering
  \includegraphics[width = 5in, bb = 0 -1 584 553] {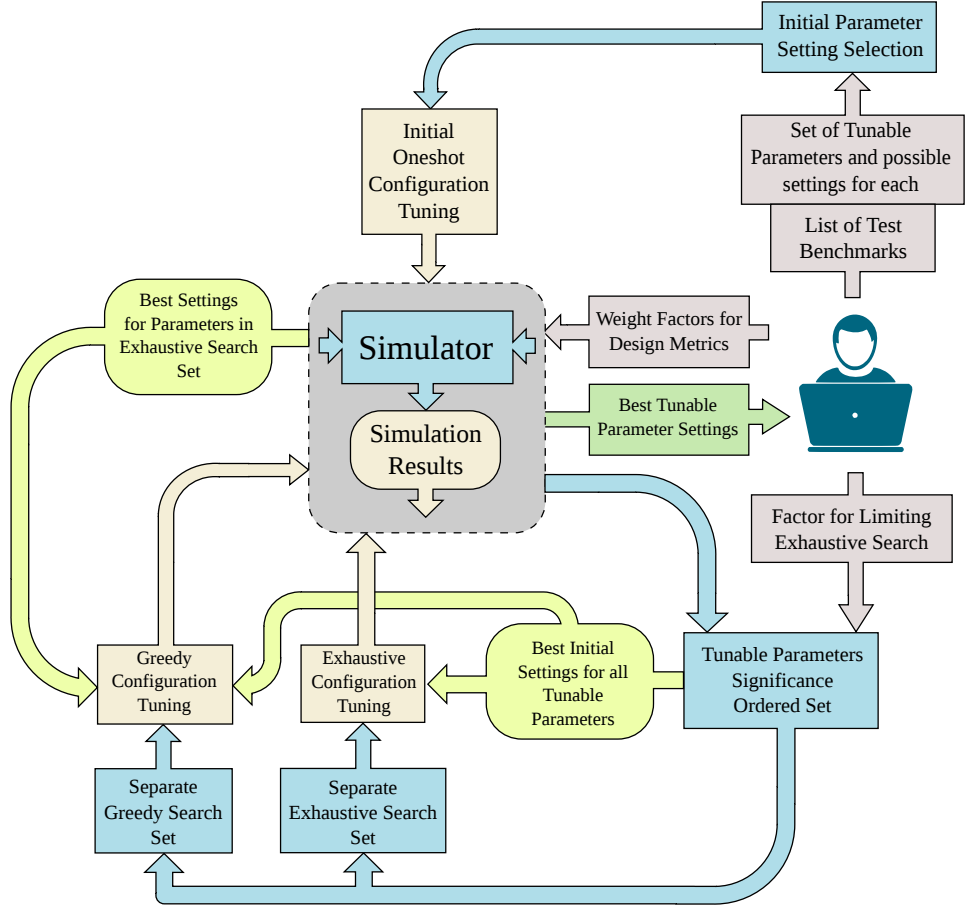}
  	\caption{Parameter Optimization Methodology.}
    \vspace{-3mm}
  	\label{figure_methodology}
\end{figure*}

Figure \ref{figure_methodology} depicts our parameter optimization methodology. The list of tunable design parameters and the set of settings for each of these parameters is provided by the system designer. The system designer also specifies the list of test benchmarks to the system. Each test benchmark provides a unique workload to the system and the system reports unique results for each benchmark. All the provided values are used by the \textit{initial parameter setting selection} module to select the initial (one-shot) configuration. The initial (one-shot) configurations are passed to the simulator module. The simulator module consists of a cycle accurate multicore architecture simulator. The simulator simulates the initial (one-shot) configurations and the obtained simulation results are normalized and combined to form an objective function using weights for different metrics specified by the system designer. The set of objective function values for each initial (one-shot) configuration is then forwarded to the \textit{tunable parameter significance ordered set} module.

In the \textit{tunable parameter significance ordered set} module, the significance of each of the tunable parameters is determined. The significance is calculated using the values of the objective function obtained from simulating the initial (one-shot) configurations. The parameters are then arranged in descending order based on their significance.

The set of ordered parameters is partitioned into three subsets. The subset with highly significant parameters is separated as the exhaustive search set, the subset with slightly less significant parameters is separated as the greedy search set followed by the subset with least significant parameters which is separated as the one-shot search set. The partition is carried out on the basis of an exhaustive search threshold factor $T$ which is used to limit the number of parameters included in the exhaustive search set. This factor is provided by the system designer. The exhaustive and greedy search sets are then forwarded to the \textit{simulator} module for use in exhaustive and greedy search steps. No further operations are carried out on the one-shot search set and the best settings obtained from the \textit{initial parameter setting selection} module are retained.

Next, exhaustive search is carried out on the exhaustive search set by the \textit{exhaustive configuration tuning} module. Although exhaustive search is a resource-intensive process, we have included this search method because it will decrease the likelihood of the solution presented by our methodology from converging to a local best instead of a global best. In this step, each possible combination of parameters in the exhaustive search set is used to generate configuration test points. For the parameters which are not in the exhaustive search set, the best settings available from the \textit{initial parameter setting selection} module are used. The objective function value for each of the exhaustive configuration test points is evaluated and the configuration yielding the smallest value is returned as the best setting for the exhaustive configuration.

The best settings reported by \textit{exhaustive configuration tuning} module are then passed on to the \textit{greedy configuration tuning} module. In this module, the parameters in the greedy search set are optimized. New configuration test points are generated by this module using parameter settings picked by a greedy algorithm for the parameters in the greedy search set. The best settings obtained from exhaustive search and the best settings obtained from initial (one-shot) optimization are used for the parameters not in the greedy search set. The greedy search algorithm operates on one parameter at a time and optimizes its settings as long as the objective function value improves for new test points. When the objective function begins to degrade, the greedy search algorithm is terminated for the current parameter and the next parameter in greedy search set is optimized. Once all the parameters in the greedy search set are optimized, the best settings for all the tunable parameters are returned to the system designer.
%
%
\subsection{Defining the Design Space}
\label{methodology_defining_the_design_space}
Consider $n$ number of tunable parameters are available to describe the design of a multicore/manycore processor. Let $P$ be the list of these tunable parameters defined as the following set:
\begin{equation}
	P = \{P_{1}, P_{2}, P_{3}, \cdots , P_{n}\}
\end{equation}
Each tunable parameter $P_{i}$ [where $i \in \{1, 2 \cdots n\}$] in the list $P$ is the set of possible settings for $i^{th}$ parameter. Let $L$ be the set containing the size of the set of possible settings for each parameter in list $P$.
\begin{equation}
	L = \{L_{1}, L_{2}, L_{3}, \cdots , L_{n}\}
\end{equation}
such that,
\begin{equation}
	L_{i} = |P_{i}| \;\; \forall \; i \in {1, 2, \cdots , n}
\end{equation}
where $|P_{i}|$ is the cardinal value of set $P_{i}$.

So, each parameter setting set $P_{i}$ in the list $P$ is defined as follows:
\begin{equation}
	P_{i} = \{P_{i1}, P_{i2}, P_{i3}, \cdots , P_{iL_{i}}\} \;\; \forall \; i \in \{1, 2, \cdots, n\}
\end{equation}
The values in the set $P_{i}$ are arranged in ascending order.

The state space for design space exploration is the collection of all the possible configurations that can be obtained using the $n$ parameters.
\begin{equation}
	S = P_{1} \times P_{2} \times P_{3} \times \cdots \times P_{n}
\end{equation}
Here, $\times$ represents the Cartesian product of lists in $P$. Throughout this paper, we use the term $S$ to denote the state space composed of all $n$ tunable parameters. To maintain generality, when referring to a state space composed of $a$ tunable parameters where $a < n$, we attach a subscript to the term $S$.
\begin{equation}
	S_{a} = P_{1} \times P_{2} \times P_{3} \times \cdots \times P_{a} \;\; \forall \; a < n
\end{equation}
We note that the state space of $a$ tunable parameters does not constitute a complete design configuration and is only used as an intermediate when defining our methodology.

We also reserve the use of $\times$ operator in the following manner:
\begin{equation}
	S_{a} = S_{a} \times P_{i} \;\; \forall \; i \in \{1, 2, \cdots, n\}
\end{equation}
This represents the extension of the state space $S_{a}$ to include one new set of parameter settings $P_{i}$ from the list $P$. This operation increases the number of tunable parameters in state space $a$ by one.

When referring to a design configuration that belongs to the state space $S$, we use the term $s$. We attach subscripts to $s$ to refer to specific design configurations. For example, a state $s_{f}$ that consists of the first setting of each tunable parameter can be written as:
\begin{equation}
	s_{f} = (P_{11}, P_{21}, P_{31}, \cdots, P_{n1})
\end{equation}
Similarly, to denote an incomplete/partial design configuration of $a$ tunable parameters we use the term $\delta s_{a}$.
%
%
\subsection{Benchmarks}
\label{methodology_benchmarks}
Each of the configurations, selected from the state space $S$ by our algorithm, is tested on $m$ number of test benchmarks. The design metrics for each simulated configuration is collected separately for each benchmark.
%
%
\subsection{Objective Function}
\label{methodology_objective_function}
In our algorithm, design configurations are compared with each other based on their objective functions. The objective function of a design configuration is the weighted sum of the design metrics obtained after simulating that design configuration. Let $o$ be the number of design metrics and $V$ be the set of values of design metrics which are obtained from the simulation.
\begin{equation}
	V_{s}^k = \{V_{s1}^k, V_{s2}^k, V_{s3}^k, \cdots , V_{so}^k\} \;\; \forall \; k = {1, 2, \cdots , m}
\end{equation}
Let $w$ be the set of weights for the design metrics based on the requirements of the targeted application. These weights are set by the system designer.
\begin{equation}
	w = \{w_{1}, w_{2}, w_{3}, \cdots , w_{o}\}\
\end{equation}
such that,
\begin{equation}
	0 \leq w_{l} \leq 1\ \forall \; l = {1, 2, \cdots , o}
\end{equation}
and,
\begin{equation}
	\sum w_{l} = 1 \;\; \forall \; l = {1, 2, \cdots , o}
\end{equation}
The objective function $\mathcal{F}$ of a design configuration $s$ for a test benchmark $k$ is defined as follows:
\begin{equation}
	\mathcal{F}_{s}^k = \sum w_{l}V_{sl}^k \;\; \forall \; l = {1, 2, \cdots , o}
\end{equation}
The optimization problem, considered in this paper, is to minimize the value of the objective function $\mathcal{F}$. The design metrics are chosen such that the minimization of their values is the favorable design choice. For example, when considering the performance metric, the design goal is to maximize performance. To model this into the objective function which we use execution time to measure performance. Minimizing execution time would fit with minimizing the objective function while still modelling the design goal of maximizing performance. The optimization problem for each test benchmark $k$ is defined as follows:
\begin{equation}
\begin{split}
	min. \;\; & F_{s}^k \\
	s.t.\;\; & s \in S
\end{split}
\end{equation}
%
%
\section{Algorithm for Parameter Optimization Methodology}
\label{algorithm}
In this section, we describe the four distinct phases that comprises our design space exploration methodology. We also compute the overall computational complexity of our methodology using the complexities of each phase.
%
%
\subsection{Phase I : Initial One-Shot Optimization and Parameter Significance}
\label{algorithm_phaseI}
In this phase, parameter tuning is carried out using one-shot optimization process. This process is based on single factor analysis which is a common approach used in design space exploration. Single factor analysis based approach is suitable for independent on/off-type parameters which have two options for each parameter, a zero value option and a non-zero value option \cite{dsDSEUDOE11}. But, from our experiment we found that when the results from single factor analysis are extended with other search algorithms (e.g. exhaustive and greedy search in our methodology) it can form an effective heuristic. The process operates on the list of parameters $P$ such that the parameters are tuned one at a time; the settings of the current parameter are varied while the settings of all other parameters are unchanged. While operating on the current parameter, objective functions are calculated for design configurations that use the first and last settings in the set of parameter settings for the current parameter. All the other parameters in these design configurations are arbitrarily set to the first setting in their corresponding set of parameter settings. Note that each set of parameters in the list of parameters $P$ is provided arranged in ascending order.

\begin{algorithm}
\caption{Initial One-Shot Optimization and Parameter Significance}
\label{algorithm_parameter_significance}

\LinesNumbered
\DontPrintSemicolon
\SetKwData{Left}{left}\SetKwData{This}{this}\SetKwData{Up}{up}
\SetKwFunction{Union}{Union}\SetKwFunction{FindCompress}{FindCompress}
\SetKwInOut{Input}{input}\SetKwInOut{Output}{output}
\KwIn{$P$ - List of Tunable Parameters}
\KwOut{$B$ - Set of Best Settings; $D$ - Significance of Parameters with respect to Objective Function}

\BlankLine
\For {$i \leftarrow 1$ \KwTo $n$}
    {
    $s_{f} = \{P_{i1}\}$\;
    $s_{l} = \{P_{iL[i]}\}$\;
    \For {$j \leftarrow 1$ \KwTo $n$}
    {
    \If {$i \neq j$}
        {
        $s_{f} = s_{f} \cup \{P_{j1}\}$\;
        $s_{l} = s_{l} \cup \{P_{j1}\}$\;
        }
    }
    \For {$k \leftarrow 1$ \KwTo $m$}
    {
    Explore $k^{th}$ benchmark using configuration $s_{f}$\;
    \ \ \ \ Calculate $\mathcal{F}_{s_{f}}^{k}$\;
    Explore $k^{th}$ benchmark using configuration $s_{l}$\;
    \ \ \ \ Calculate $\mathcal{F}_{s_{l}}^{k}$\;
    $D^{k}_{i}$ = $\mathcal{F}^{k}_{l} - \mathcal{F}^{k}_{f}$\;
    \label{algorithm_parameter_significance_ref1}
    \eIf {$D^{k}_{i} > 0$}
        {
        $B^{k}_{i} = P_{i1}$\;
        \label{algorithm_parameter_significance_ref2}
        }
        {
        $B^{k}_{i} = P_{iL[i]}$\;
        \label{algorithm_parameter_significance_ref3}
        }
    }
}
\end{algorithm}
Algorithm \ref{algorithm_parameter_significance} presents the steps in selecting initial parameter setting of each tunable parameter, and determining the significance of each tunable parameter to the objective function. In this algorithm, two design configurations, $s_{f}$ and $s_{l}$, are used where $s_{f}$ corresponds to the design configuration with the first setting and $s_{l}$ corresponds to the design configuration with the last setting for the current parameter being processed. The objective functions for all the test benchmarks are calculated by running simulations on these design configurations. The objective functions $\mathcal{F}_{s_{f}}$ and $\mathcal{F}_{s_{l}}$, which  corresponds to design configurations $s_{f}$ and $s_{l}$, respectively, are compared with each other. The comparison is made on the basis of difference of magnitude of $\mathcal{F}_{s_{f}}$ from $\mathcal{F}_{s_{l}}$. This difference is stored in a set of parameter significance $D$ (line \ref{algorithm_parameter_significance_ref1}). The magnitude of the difference determines the significance of parameters with respect to objective function, that is, greater the magnitude $D_{i}^{k}, \; i \in \{1, 2, 3, \ldots, n\}$, the greater the significance of parameter $P_{i}$ with respect to the objective function. Judging by whether the difference is positive or negative, the best setting for the current parameter is chosen as either the first setting or the last setting. The best settings for the parameters are stored in the set of best settings $B_{i}^{k}$ (lines \ref{algorithm_parameter_significance_ref2} and \ref{algorithm_parameter_significance_ref3}).
%
%
\subsection{Phase II : Set Partitioning}
\label{algorithm_phaseII}
\begin{algorithm}
\caption{Set Partitioning}
\label{algorithm_set_partitioning}

\LinesNumbered
\DontPrintSemicolon
\SetKwData{Left}{left}\SetKwData{This}{this}\SetKwData{Up}{up}
\SetKwFunction{Union}{Union}\SetKwFunction{FindCompress}{FindCompress}
\SetKwInOut{Input}{input}\SetKwInOut{Output}{output}
\KwIn{$D$ - Significance of Parameters towards Objective Function; $I$ - Index Set; $T$ - Exhaustive Search Threshold Factor}
\KwOut{$\mathcal{E}$ - Set of Parameters for Exhaustive Search; $\mathcal{G}$ - Set of Parameters for Greedy Search}
\BlankLine
$\mathcal{E} = \emptyset\ and\ \mathcal{G} = \emptyset$\;
\For {$k \leftarrow 1$ \KwTo $m$}
    {
    sortDescending ($\mid D^{k} \mid$)- s.t. index information of the sorted values is preserved in $I^{k}$\;
    \label{algorithm_set_partitioning_ref1}
    sort($P^{k}$) and sort($L^{k}$) w.r.t. index information in $I^{k}$\;
    $num_{\mathcal{E}} = 1$ and $i = 1$\;
    \While {$num_{\mathcal{E}} \leq T$}
        {
		\label{algorithm_set_partitioning_ref2}
        $num_{\mathcal{E}} = num_{\mathcal{E}} \times L_{i}^{k}$\;
	    \eIf {$num_{\mathcal{E}} \leq T$}
	    {
           $\mathcal{E}^{k} = \mathcal{E}^{k} \cup \{P_{i}\}$\;
           $i = i + 1$\;
	    }
        {
	      \textbf{break}\;
        }
        }
    $num_{\mathcal{G}} = ceil((|P^{k}| - |\mathcal{E}^{k}|)\ /\ 2)$\;
    \label{algorithm_set_partitioning_ref3}
    \While {$num_{\mathcal{G}} > 0$}
        {
        $\mathcal{G}^{k} = \mathcal{G}^{k} \cup \{P_{i}^{k}\}$\;
        $num_{\mathcal{G}} = num_{\mathcal{G}} - 1$\;
        $i = i + 1$\;
        }
    }
\end{algorithm}
Algorithm \ref{algorithm_set_partitioning} presents the steps involved in partitioning the list of parameters into different sets for performing the exhaustive and greedy searches. The algorithm starts out by sorting the set of parameter significance values, $\mid D^{k} \mid$, in descending order of magnitude such that the index information of the sorted values is preserved in $I^{k}$ (for the $k$th benchmark; $k \in \{1, 2, \ldots, m\}$) (line \ref{algorithm_set_partitioning_ref1}). For example, if the fifth entry $D_{5}^{k}$ has the greatest value, $D_{5}$ will become the first entry after sortDescending($\mid D^{k} \mid$) function and first entry of the set $I^{k}$ will be 5, that is, $I_{1}^{k} = 5$. The index information in $I^{k}$ is then used by the sort($P^{k}$) and sort ($L^{k}$) functions to order the list of parameter names and set sizes. This results in an arrangement of parameters in which the parameters with higher significance are placed towards the start of the set and the parameters with lower significance are placed towards the end of the set. The parameters are then sorted into three ordered sets. The parameters with the highest significance in the sorted parameters list are separated into the exhaustive search set $\mathcal{E}$. The number of parameters considered for the exhaustive search set depends upon the exhaustive search threshold factor $T$, provided by the system designer. The threshold factor $T$ limits the size of the partial search space of the exhaustive search set, $num_{\mathcal{E}}$ (line \ref{algorithm_set_partitioning_ref2}).

When the exhaustive search set is separated out, the remaining parameters in the sorted parameters list are halved (line \ref{algorithm_set_partitioning_ref3}). The upper half of remaining parameters in the sorted parameters list are separated into the greedy search set, $\mathcal{G}$, and the lower half is separated into one-shot search set. We observe empirically that selecting the upper half, $ceil((|P| - |\mathcal{E}^{k}|)/2)$, provides efficient design space exploration without significantly compromising the solution (i.e., best design configuration) quality. The parameters in the one-shot search set are not explored further and are tuned using the best settings determined for them in Algorithm \ref{algorithm_parameter_significance}.
%
%
\subsection{Phase III : Exhaustive Search Optimization}
\label{algorithm_phaseIII}
\begin{algorithm}
\caption{Exhaustive Search}
\label{algorithm_exhaustive_search}

\LinesNumbered
\DontPrintSemicolon
\SetKwData{Left}{left}\SetKwData{This}{this}\SetKwData{Up}{up}
\SetKwFunction{Union}{Union}\SetKwFunction{FindCompress}{FindCompress}
\SetKwInOut{Input}{input}\SetKwInOut{Output}{output}
\KwIn{$P$ - List of Tunable Parameters; $B$ - Set of Best Settings for Oneshot Search; $\mathcal{E}$ - List of Parameters for Exhaustive Search}
\KwOut{$B$ - List of Best Settings for Oneshot and Exhaustive Search}
\BlankLine
$s_{\mathcal{E}} = \emptyset$\;
$\delta s_{\mathcal{E}} = \emptyset\ and\ \delta s_{\mathcal{E}^{'}} = \emptyset$\;

\For {$k \leftarrow 1$ \KwTo $m$}
    {
    $\mathcal{F}^{k}_{s_{b}} = \infty$\;
    \For {$i \leftarrow 1$ \KwTo $n$}
        {
        \If {$P_{i} \notin \mathcal{E}^{k}$}
        		{
	        $\delta s^{k}_{\mathcal{E}^{'}} = \delta s^{k}_{\mathcal{E}^{'}} \cup \{B^{k}_{i}\}$\;	
	        \label{algorithm_exhaustive_search_ref1}
        		}
        }

    \For {$i \leftarrow 1$ \KwTo $n$}
        {
        \If {$P_{i} \in \mathcal{E}^{k}$}
	        {
        		$S^{k}_{\mathcal{E}} = S^{k}_{\mathcal{E}} \times P_{i}$\;
        		\label{algorithm_exhaustive_search_ref2}
        		}
        	}

    \For {$j \leftarrow 1$ \KwTo $|S^{k}_{\mathcal{E}}|$}
        {
        	$\delta s^{k}_{\mathcal{E}j}$ is a partial configuration in state space $S^{k}_{\mathcal{E}}$\;
        	\label{algorithm_exhaustive_search_ref3}
        $s^{k}_{\mathcal{E}} = \delta s^{k}_{\mathcal{E}j} \cup \delta s^{k}_{\mathcal{E}^{'}}$\;
        \label{algorithm_exhaustive_search_ref4}
        Explore $k^{th}$ benchmark using configuration $s^{k}_{\mathcal{E}}$\;
        \ \ \ \ Calculate $\mathcal{F}^{k}_{\mathcal{E}}$\;
        \If {$\mathcal{F}^{k}_{s_{\mathcal{E}}} < \mathcal{F}^{k}_{s_{b}}$}
            {
            \label{algorithm_exhaustive_search_ref5}
            $\mathcal{F}^{k}_{s_{b}} = \mathcal{F}^{k}_{s_{\mathcal{E}}}$\;
            $B^{k} = s^{k}_{\mathcal{E}}$\;
            }
        }
    }
\end{algorithm}
In the third phase of our methodology, best settings for the parameters in the exhaustive search set $\mathcal{E}$ are determined. The steps involved in this phase are described in Algorithm \ref{algorithm_exhaustive_search}. First, the settings for the parameters that are not included in the set $\mathcal{E}$ are chosen (line \ref{algorithm_exhaustive_search_ref1}). The settings for these parameters are assigned using the set of best settings $B_{i}^{k}$ obtained from the initial (one-shot) optimization process in Algorithm \ref{algorithm_parameter_significance}. These settings are stored in the set $\delta s_{\mathcal{E}^{'}}$. Following this, a partial state space $S_{\mathcal{E}}$ for exhaustive search set $\mathcal{E}$ is formed (line \ref{algorithm_exhaustive_search_ref2}). Next, each of the possible configurations of parameter settings, $\delta s_{\mathcal{E}j}$ (line \ref{algorithm_exhaustive_search_ref3}), in the state-space of set $\mathcal{E}$, $S_{\mathcal{E}}$, is combined with the set $\delta s_{\mathcal{E}^{'}}$ (line \ref{algorithm_exhaustive_search_ref4}) to form a complete simulatable design configuration $s_{\mathcal{E}}$. The objective function, $\mathcal{F}_{s_{\mathcal{E}}}$, is determined by simulating $s_{\mathcal{E}}$, and this value is compared with the best objective function, $\mathcal{F}_{s_{b}}$, which holds the smallest value objective function encountered thus far in the search process. When a design configuration results in an objective function that has a value less than $\mathcal{F}_{s_{b}}$ (line \ref{algorithm_exhaustive_search_ref5}), then, $\mathcal{F}_{s_{b}}$ is changed to the new minimum value and the set of best settings $B$ is updated with the corresponding design configuration. This exhaustive search optimization process is carried out for each of the test benchmarks used in the simulation process.
%
%
\subsection{Phase IV : Greedy Search Optimization}
\label{algorithm_phaseIV}
Algorithm \ref{algorithm_greedy_search}, describes the final phase of our methodology which involves optimization of the parameters in greedy search set $\mathcal{G}$. For each parameter in the set $\mathcal{G}$, the significance of that parameter is checked to determined whether the value is positive or negative. A positive value indicates that the first setting for that parameter yields a smaller objective function as compared to the last setting. A negative value indicates the exact opposite; the last setting for that parameter yields a smaller objective function as compared to the first setting. It is assumed that the setting that yields the smallest objective function lies closer towards the setting that yields the smallest objective function in the initial (one-shot) optimization process. So, to ensure that the search process starts from the setting that yielded the smallest objective function in the initial (one-shot) optimization process, the set of parameter settings is either sorted in descending order or left unchanged in default ascending order (line \ref{algorithm_greedy_search_ref1}).
\begin{algorithm}
\caption{Greedy Search}
\label{algorithm_greedy_search}

\LinesNumbered
\DontPrintSemicolon
\SetKwData{Left}{left}\SetKwData{This}{this}\SetKwData{Up}{up}
\SetKwFunction{Union}{Union}\SetKwFunction{FindCompress}{FindCompress}
\SetKwInOut{Input}{input}\SetKwInOut{Output}{output}
\KwIn{
$P$ - List of Tunable Parameters, $D$ - Significance of Parameters towards Objective Function, $B$ - Set of Best Settings for Oneshot and Exhaustive Search, $\mathcal{E}$ - Set of Parameters for Exhaustive Search, $\mathcal{G}$ - Set of Parameters for Greedy Search}
\KwOut{$B$ - Complete set of Best Settings}
\BlankLine
$s_{\mathcal{G}} = \emptyset$\;
$\delta s_{\mathcal{G}^{'}} = \emptyset$\;
$\mathcal{G}_{P} = \emptyset$\;
\For {$k = 1$ \KwTo $m$}
    {
    $\mathcal{F}^{k}_{s_{b}} = \infty$\;
    \For {$i \leftarrow 1$ \KwTo $n$}
        {
        \If {$P_{i} \in \mathcal{G}^{k}$}
            {
            \If {$D^{k}_{i} < 0$}
                {
                \label{algorithm_greedy_search_ref1}
                $\mathcal{G}_{P}$ = sortDescending ($P_{i}$)\;
                }
            \For {$j \leftarrow 1$ \KwTo $n$}
                {
                \If { $P_{j} \neq \mathcal{G}_{P}$}
                    {
                    \label{algorithm_greedy_search_ref2}
                    $\delta s^{k}_{\mathcal{G}_{P}^{'}} = \delta s^{k}_{\mathcal{G}_{P}^{'}} \cup \{B^{k}_{j}\}$\;
                    }
                }
            \For {$l \leftarrow 1$ \KwTo $L_{i}$}
                {
                $s^{k}_{\mathcal{G}} = \delta s^{k}_{\mathcal{G_{P}}^{'}} \cup \{\mathcal{G}_{Pl}\}$\;
                \label{algorithm_greedy_search_ref3}
                Explore $k^{th}$ benchmark using configuration $s^{k}_{\mathcal{G}}$\;
                \ \ \ \ Calculate $\mathcal{F}^{k}_{s_{\mathcal{G}}}$\;
                \eIf {$\mathcal{F}^{k}_{s_{\mathcal{G}}} < \mathcal{F}^{k}_{s_{b}}$}
                    {
                    \label{algorithm_greedy_search_ref4}
                    $\mathcal{F}^{k}_{s_{b}} = \mathcal{F}^{k}_{s_{\mathcal{G}}}$\;
                    $B^{k}_{i} = \mathcal{G}_{Pj}$\;
                    }
                    {
                    \textbf{break}\;
                    }
                }
            }
        }
    }
\end{algorithm}
In the search process of each parameter, all the parameters except for the parameter currently being processed is assigned settings using the set of best settings $B_{i}^{k}$ (line \ref{algorithm_greedy_search_ref2}). The parameters in the exhaustive search set $\mathcal{E}$ are set to the best settings obtained from the exhaustive search process. The non-current parameters in set $\mathcal{G}$ are assigned settings in one of two ways. If the non-current parameter has already been processed by the greedy search optimization process, then, the parameter is assigned the best setting obtained from that process. If the non-current parameter has not been processed yet, then, the parameter is assigned the best setting obtained from the initial (one-shot) optimization process. The parameters that are not included in either of the sets $\mathcal{E}$ or $\mathcal{G}$ are assigned the best settings obtained from the initial one-shot optimization process. These settings are collected in a partial design configuration set $\delta s_{\mathcal{G_{P}}^{'}}$.

The set $\delta s_{\mathcal{G_{P}}^{'}}$ then is combined with the settings for the current parameter being processed to form the complete simulatable design configuration $s_{\mathcal{G}}$ (line \ref{algorithm_greedy_search_ref3}). This configuration is simulated and the resulting objective function, $\mathcal{F}_{s_{\mathcal{G}}}$, is compared with the best objective function $\mathcal{F}_{s_{b}}$,  which holds the smallest value objective function encountered thus far in the search process. Similar to the exhaustive search process, when a design configuration results in an objective function that has a value less than $\mathcal{F}_{s_{b}}$ (line \ref{algorithm_greedy_search_ref4}), then, $\mathcal{F}_{s_{b}}$ is changed to the new minimum value and the set of best settings $B_{i}^{k}$ is updated with the corresponding design configuration. However, when the search process encounters a design configuration that results in an objective function that has a value greater than $\mathcal{F}_{s_{b}}$, then, the search process for the current parameter is terminated and the next parameter in the parameter list $\mathcal{G}$ is explored.

\subsection{Computational Complexity}
\label{algorithm_complexity}
The computational complexity for our design space exploration methodology is $\mathcal{O}(mT + mnL_{max}$log$L_{max})$. This term is comprised of the initial one-shot parameter optimization and parameter significance (Algorithm~\ref{algorithm_parameter_significance}) $\mathcal{O}(mn)$, set partitioning (Algorithm~\ref{algorithm_set_partitioning}) $\mathcal{O}(mn$log$n)$ (sorting contributes the $n$log$n$ factor), exhaustive search (Algorithm~\ref{algorithm_exhaustive_search}) $\mathcal{O}(mT)$, and greedy search (Algorithm~\ref{algorithm_greedy_search}) $\mathcal{O}(mnL_{max}$log$L_{max})$, where, $n$ is the number of tunable parameters, $m$ is the number of test benchmarks, $T$ is the exhaustive search threshold factor and $L_{max}$ is the size of the largest set of possible settings for the tunable parameters considered. Since $T$ is larger than $n$, $m$ and $L_{max}$, the computational complexity of our methodology can be simplified as $\mathcal{O}(mT)$. This complexity reveals that the operation time of our proposed methodology depends on the exhaustive search threshold factor provided by the system designer.
%
%
\section{Experimental Setup}
\label{experimental_setup}
All the design configuration simulations for evaluating our parameter optimization algorithm were carried out on the ESESC \cite{ekaESESC13} simulator. ESESC is a fast cycle-accurate chip multiprocessor simulator which models out-of-order RISC (Reduced Instruction Set Computing) processors running ARM instruction set. It produces performance, power and thermal models of chip multi-processors. It also has the ability to run unmodified ARM-binaries.

The test benchmarks used in our evaluation are from the PARSEC and SPLASH-2 \cite{ybPARSEC3_TUT11} \cite{cbPARSEC11} \cite{mgPARSEC_M509} Benchmark suite. The PARSEC and SPLASH-2 benchmark suite is a collection of standardized multi-threaded benchmarks used for evaluating chip-multiprocessors. Unlike high performance computing (HPC) focused benchmark suites, the benchmarks included in this suite are representative of a diverse application space. These benchmarks model emerging workloads which are likely to find important applications in the near future.

The algorithm steps were implemented using Perl \cite{pmPERL15}. The results from each of the simulations carried out were gathered in MS Excel using the tool Excel-Writer-XLSX \cite{jmExcel15} for Perl. The ARM-binaries for all the test benchmarks were compiled using arm-linux-gnueabihf toolchain \cite{qzBPFA12}.

\begin{table}[!h]
\caption{List of Tunable Parameters and Settings}
\label{table_list_of_parameters}
\centering
\begin{tabular}[width = \columnwidth]{|l|c|c|c|}\hline
\multirow{2}{*}{\textbf{Parameter Name}}          & \multirow{2}{*}{\textbf{Set of Settings}} & \multicolumn{2}{c|}{\textbf{Design Space}}\\\cline{3-4}
& & \textbf{Small} & \textbf{Large}\\\hline
Cores (PARSEC)               & 2, 4, 8                            & $\checkmark$ & $\checkmark$\\\hline
Cores (SPLASH-2)             & 2, 4                               & $\checkmark$ & $\checkmark$\\\hline
Frequency (MHz)              & 1700, 2200, 2800, 3200             & $\checkmark$ & $\checkmark$\\\hline
L1-I Cache Size (kB)         & 8, 16, 32, 64, 128                 & $\checkmark$ & $\checkmark$\\\hline
L1-D Cache Size (kB)         & 8, 16, 32, 64, 128                 & $\checkmark$ & $\checkmark$\\\hline
L2 Cache Size (kB)           & 256, 512, 1024                     & $\checkmark$ & $\checkmark$\\\hline
L3 Cache Size (kB)           & 2048, 4096, 8192                   & $\checkmark$ & $\checkmark$\\\hline
Fetch/Issue/Retire           & \multirow{2}{*}{2, 4, 8, 16}       &              & \multirow{2}{*}{$\checkmark$}\\
Width (B)                    &                                    &              &             \\\hline
Reorder Buffer Size (B)      & 32, 64, 128, 256                   &              & $\checkmark$\\\hline
Branch Prediction            & BPredX, BPredX2                    &              & $\checkmark$\\\hline
\end{tabular}
\end{table}

We tested our parameter optimization methodology for different design space sizes: small and large. The smaller design space contains six tunable parameters while the larger design space contains nine. Table \ref{table_list_of_parameters} contains the list of tunable parameters considered in making up the design space along with the set possible setting values for each of the parameters. Based on the number of settings for each of the tunable parameters the design space cardinalities for PARSEC benchmarks is 2,700 design points for the smaller design space and 86,400 for the larger design space. For SPLASH-2 benchmarks, the design space cardinality of the smaller design space is 1,800 and for the larger design space is 57,600.

We used the following benchmarks from the PARSEC and SPLASH-2 suites to test our algorithm:

\textbf{PARSEC Benchmarks}: Blackscholes, Canneal, Facesim, Fluidanimate, Freqmine, Streamcluster, Swaptions, x264

\textbf{SPLASH-2 Benchmarks}: Cholesky, FFT, LU-cb, LU-ncb, Ocean-cp, Ocean-ncp, Radiosity, Radix, Raytrace

\begin{figure}[!h]
  \centering
  \includegraphics[width = 0.44\textwidth, bb = 0 1 400 327] {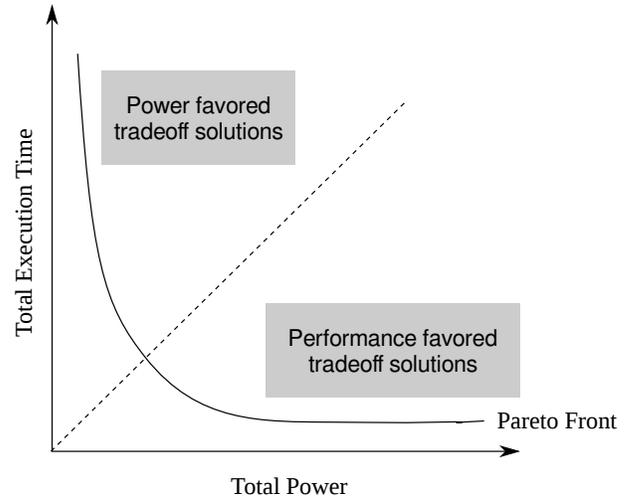}
  	\caption{Categorization of trade-off solutions in Pareto front\cite{jmUPPPALPD14}.}
  \label{figure_pareto_regions}
\end{figure}

We define two sample application domains which we use to evaluate our methodology: a low-power application domain and a high-performance application domain. In order to model the requirements of these domains, we choose power and performance as the design metrics. The total dynamic power and leakage power across all the cores of the multicore/manycore system makes up the power metric and the total execution time makes up the performance metric. The power and performance of multicore systems are known to have conflicting interdependency i.e. it is not possible to find a configuration point in the design space which yields minimum values for both power and performance. So, only trade-off solutions are possible for these metrics. The trade-off solutions for these metrics fall into two categories - low-power or high-performance (low execution time) (shown in Figure \ref{figure_pareto_regions}).

Out of the trade-off solutions, one of the solutions has to be selected for the design. Intuitively, selecting the solution at extreme minimum values of the design metrics should be ideal. This is not the case because a solution that has the lowest power value has very high execution time and the solution having lowest execution time has very high power due to conflicting interdependency of the metrics. So, the best solution is the one in which the principle design metric for the application specification is close to the minimum value while the other metrics are within acceptable ranges. This balance is introduced by applying suitable weights to the design metrics to form an objective function which quantifies the effect of each metric on the target design. The selected trade-off solution is one which yields minimum value for the objective function.

The weight assigned to the design metrics for the application domains we consider are presented in Table \ref{table_metric_weights}.
\begin{table}
\caption{Weight Factors for Design Metrics}
\label{table_metric_weights}
\centering
\begin{tabular}[width = \columnwidth]{| l | c | c |}\hline
Configuration                & Power & Performance  \\\hline
Low Power                    &  0.9  &    0.1       \\\hline
High Performance             &  0.1  &    0.9       \\\hline
\end{tabular}
\end{table}
The linear objective function we formed for our evaluation tests is the weighted sum of design metrics:
\begin{equation}
	\mathcal{F} = w_{\mathcal{P}} \cdot \mathcal{P} + w_{E} \cdot E
\end{equation}
where,
\begin{equation}
\begin{aligned}
  \mathcal{P} & = Dynamic\ Power + Leaked\ Power \\
  E & = Total\ Execution\ Time \\
\end{aligned}
\end{equation}
where, $w_{\mathcal{P}}$ and $w_{\mathcal{E}}$ denote weight factors for power and performance metrics, respectively.
%
%
\section{Results}
\label{results}
In this section, we present the results that we obtained from our methodology. This section is divided into two subsections. In the first subsection, we present the evaluation of results obtained from our methodology on the smaller design space - six parameters. We further compare our results with the results obtained from a fully exhaustive search of the smaller design space. In the second subsection, we present the evaluation of the results obtained from our methodology on the larger design space. We also compare the results using different exhaustive search threshold factors ($T$) and the effect of these threshold factors on solution quality and design time.
%
%
\subsection{Evaluation Results for Smaller Design Space}
\label{results_smaller_design_space}
\begin{figure*}[!t]
  \centering
  \subfloat[Low-power requirement]
	  {
    \includegraphics[width = 0.44\textwidth, bb = 0 0 706 536] {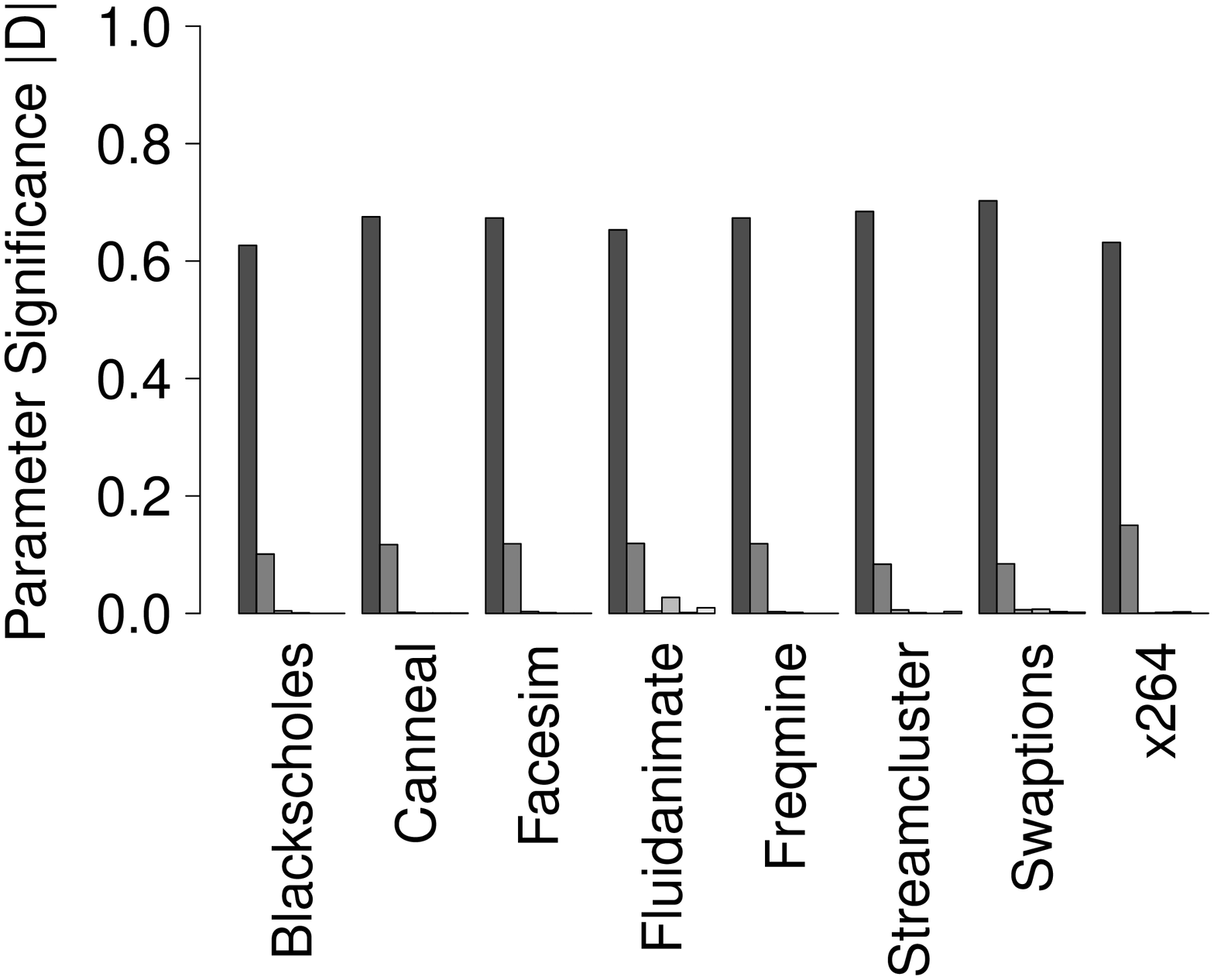}
    \label{figure_smaller_design_space_parameter_significance_low_power}
    }
  \hspace{8mm}
  \subfloat[High-performance requirement]
	  {
    \includegraphics[width = 0.44\textwidth, bb = 0 0 706 536] {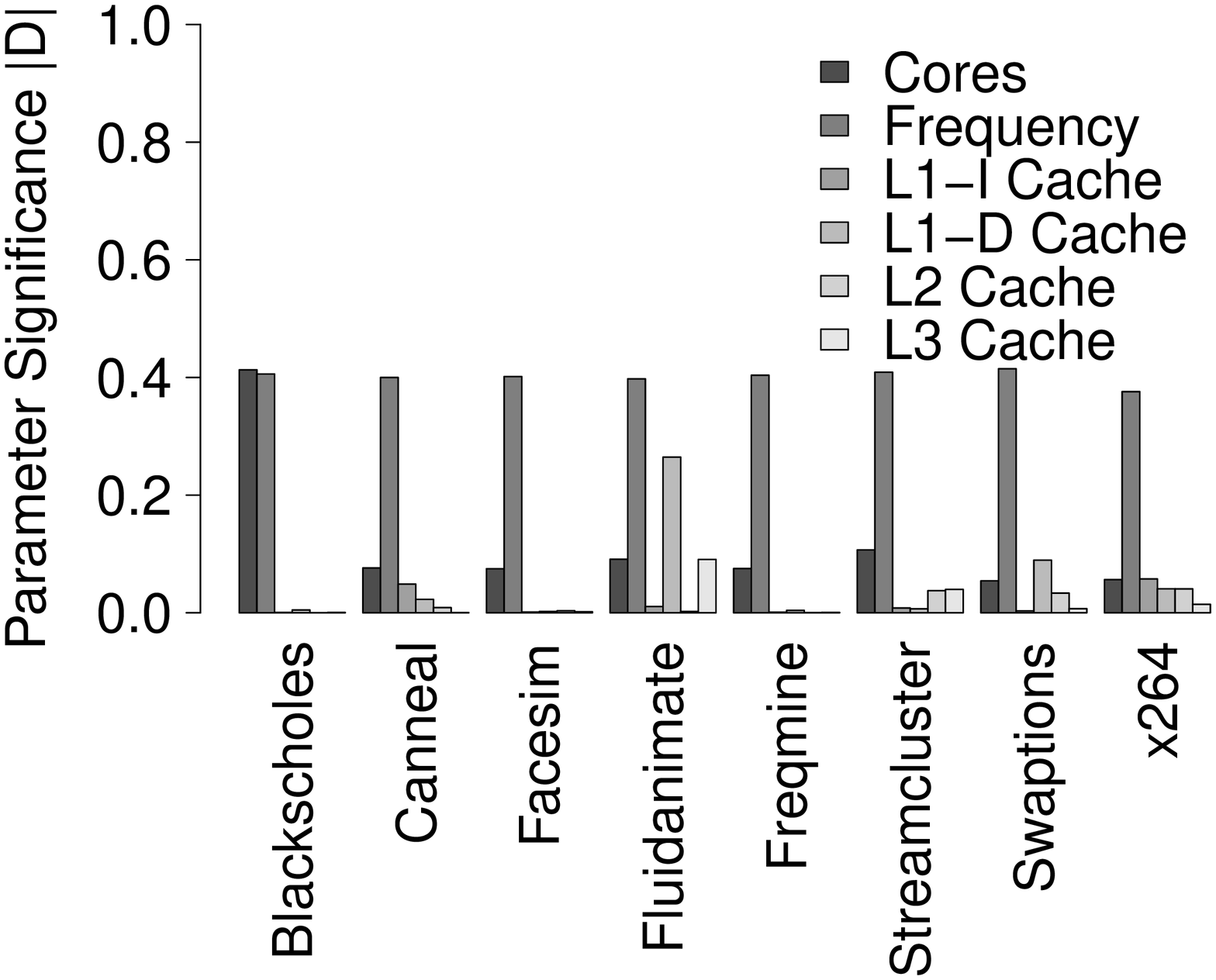}
    \label{figure_smaller_design_space_parameter_significance_high_performance}
  	  }
  \caption{Significance of Tunable parameters for PARSEC Benchmarks for small design space.}
  \vspace{-3mm}
  \label{figure_smaller_design_space_parameter_significance}
\end{figure*}

For the smaller design space, we verified our methodology by comparing our results with the results obtained from fully exhaustive search. We execute our methodology with an arbitrary exhaustive search threshold factor of 150 (i.e., $T = 150$). We clarify that the threshold factor of $T = 150$ means that the maximum number of design configurations that can be explored by exhaustive search (Algorithm \ref{algorithm_exhaustive_search}) in our methodology is upper bounded by 150.
%
%
\subsubsection{Parameter Significance}
\label{smaller_design_space_parameter_significance}
\begin{figure*}[!b]
  \centering
  \subfloat[Low-power requirement]
	  {
	  \includegraphics[width = 0.44\textwidth , bb = 0 0 706 536] {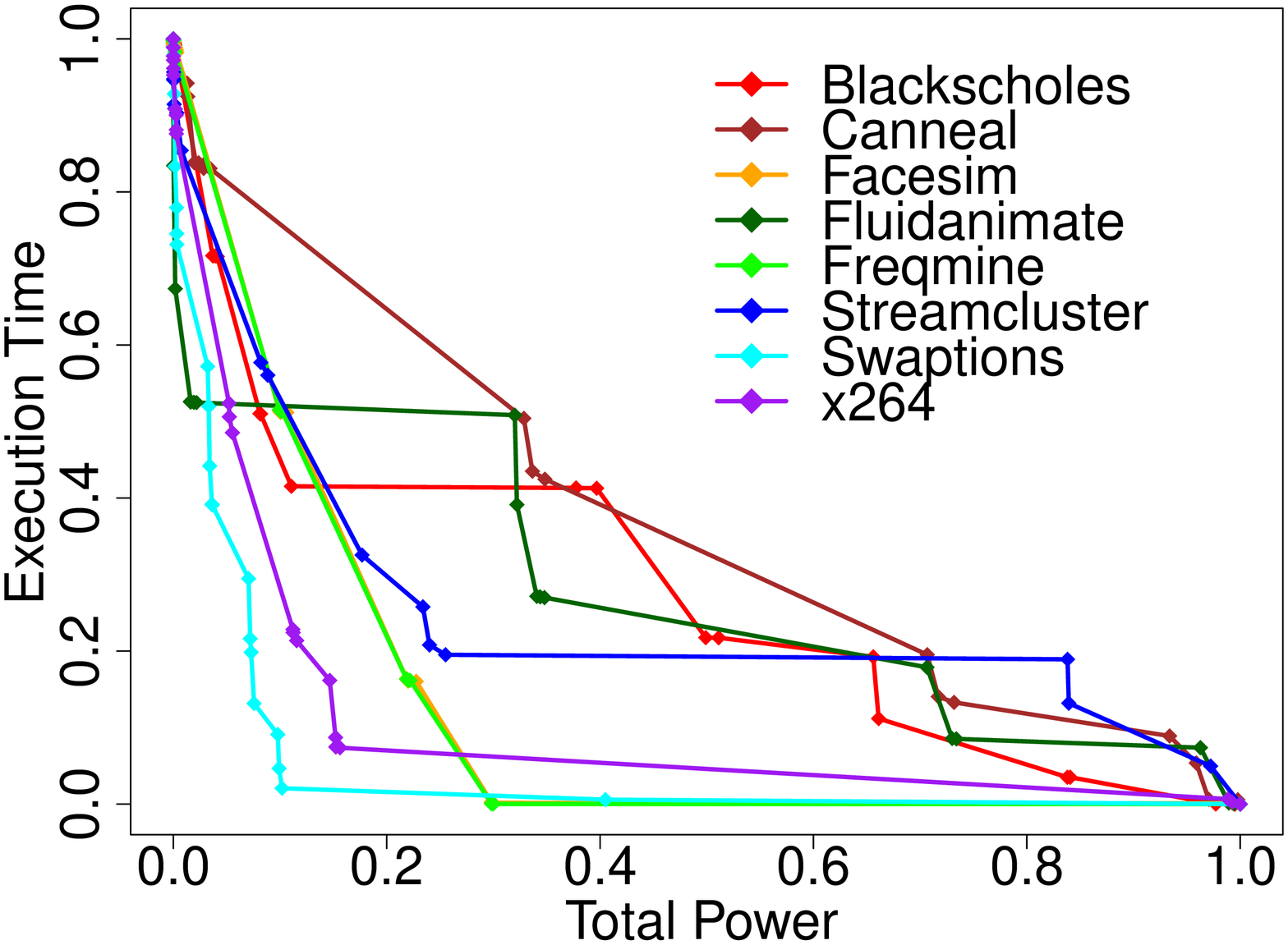}
    \label{figure_smaller_design_space_pareto_fronts_low_power}
  	  }
  \hspace{8mm}
  \subfloat[High-performance requirement]
	  {
	  \includegraphics[width = 0.44\textwidth , bb = 0 0 706 536] {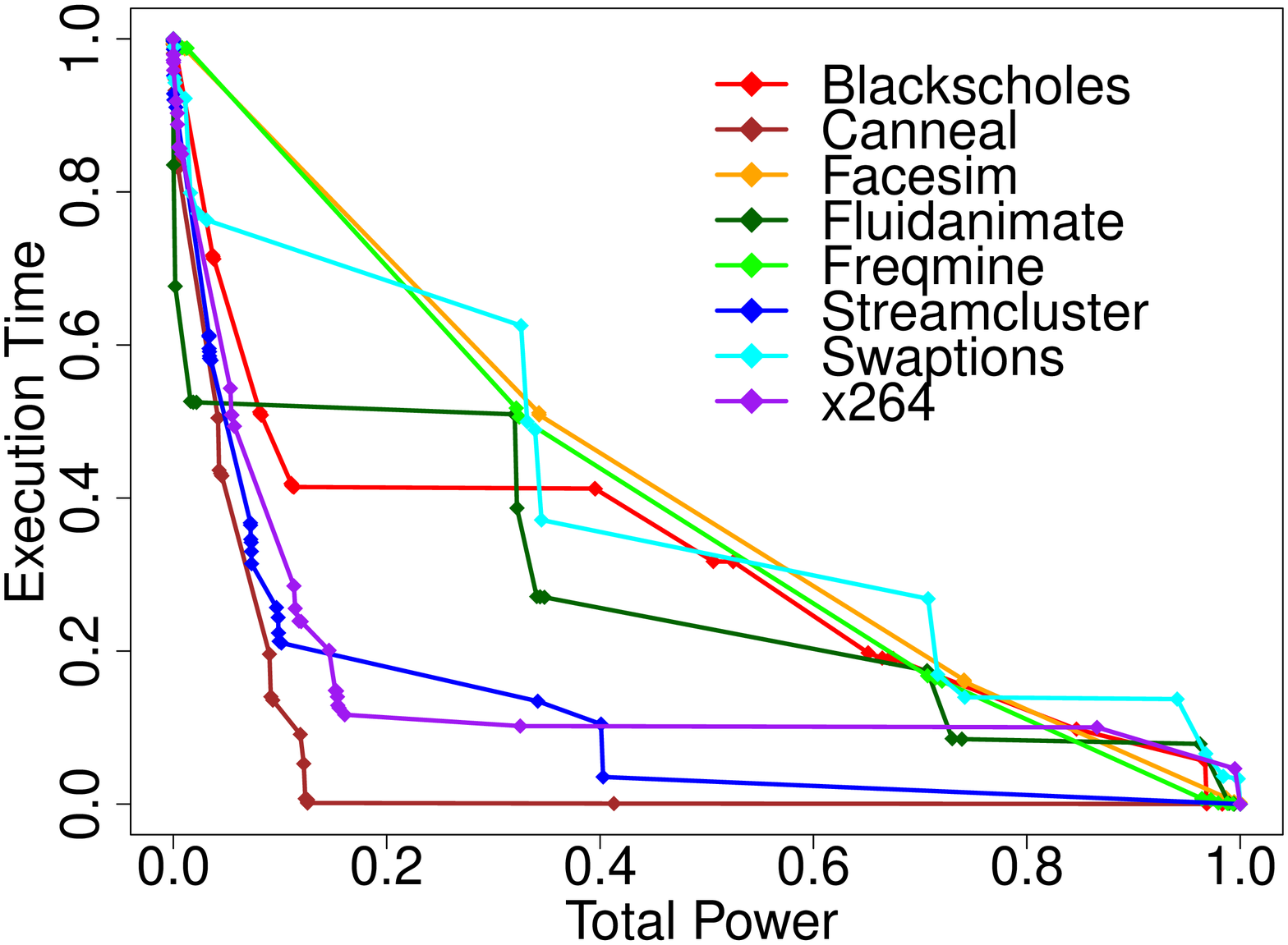}
    \label{figure_smaller_design_space_pareto_fronts_high_performance}
    	}
  \caption{Pareto fronts for PARSEC Benchmarks for small design space with exhaustive search threshold of T = 150 [total power and execution time values are normalized].}
  \vspace{-3mm}
  \label{figure_smaller_design_space_pareto_fronts}
\end{figure*}

Figure \ref{figure_smaller_design_space_parameter_significance} presents the values of parameter significance obtained for various PARSEC benchmarks for both low-power and high-performance requirements. These values were calculated based on normalized values of the simulation results obtained from the initial (one-shot) optimization phase of our methodology. Total power and execution time values are normalized using the maximum total power and maximum execution time values obtained in the initial (one-shot) optimization phase. All total power and execution time values throughout the remaining phases of the methodology are normalized using these maximum values. Figure \ref{figure_smaller_design_space_parameter_significance} reveals that for each test benchmark, there are at most three tunable parameters which have a high significance on the objective function. We observe that for low-power requirements, number of cores followed by the core frequency are the parameters that have significant impact on the objective function. For high-performance requirements, cache sizes (in particular L1-D and L2 cache sizes) also impact the objective function as the larger cache sizes can accommodate more recently used data and help mitigate cache thrashing, which provides performance improvement. This observation is intuitive because the PARSEC benchmarks are highly data-parallel (and hence performance improvement with an increase in number of cores) and have medium to large working sets (and hence performance improvement with an increase in cache size).
%
%
\subsubsection{Pareto Fronts}
\label{smaller_design_space_pareto_fronts}
Figure \ref{figure_smaller_design_space_pareto_fronts} presents the Pareto fronts obtained for various PARSEC benchmarks for both low-power and high-performance requirements. The Pareto fronts, which are generated using the normalized values of total power and execution time clearly show the conflicting interdependency between these design metrics considered, i.e., increasing total power decreases execution time and vice versa. Hence, it is not possible to have a single solution to the optimization problem which gives minimum values for both these metrics. A favorable trade-off solution between power and performance is the only result that can be obtained from the optimization process. Each of the points on the Pareto front represents a favorable trade-off solution between the conflicting power and performance metrics.
%
%
\subsubsection{Selecting a favorable tradeoff solution}
\label{smaller_design_space_tradeoff_solution}
\begin{figure*}[!t]
  \centering
  \subfloat[Low-power requirement]
	  {
    \includegraphics[width = 0.44\textwidth, bb = 0 0 706 536] {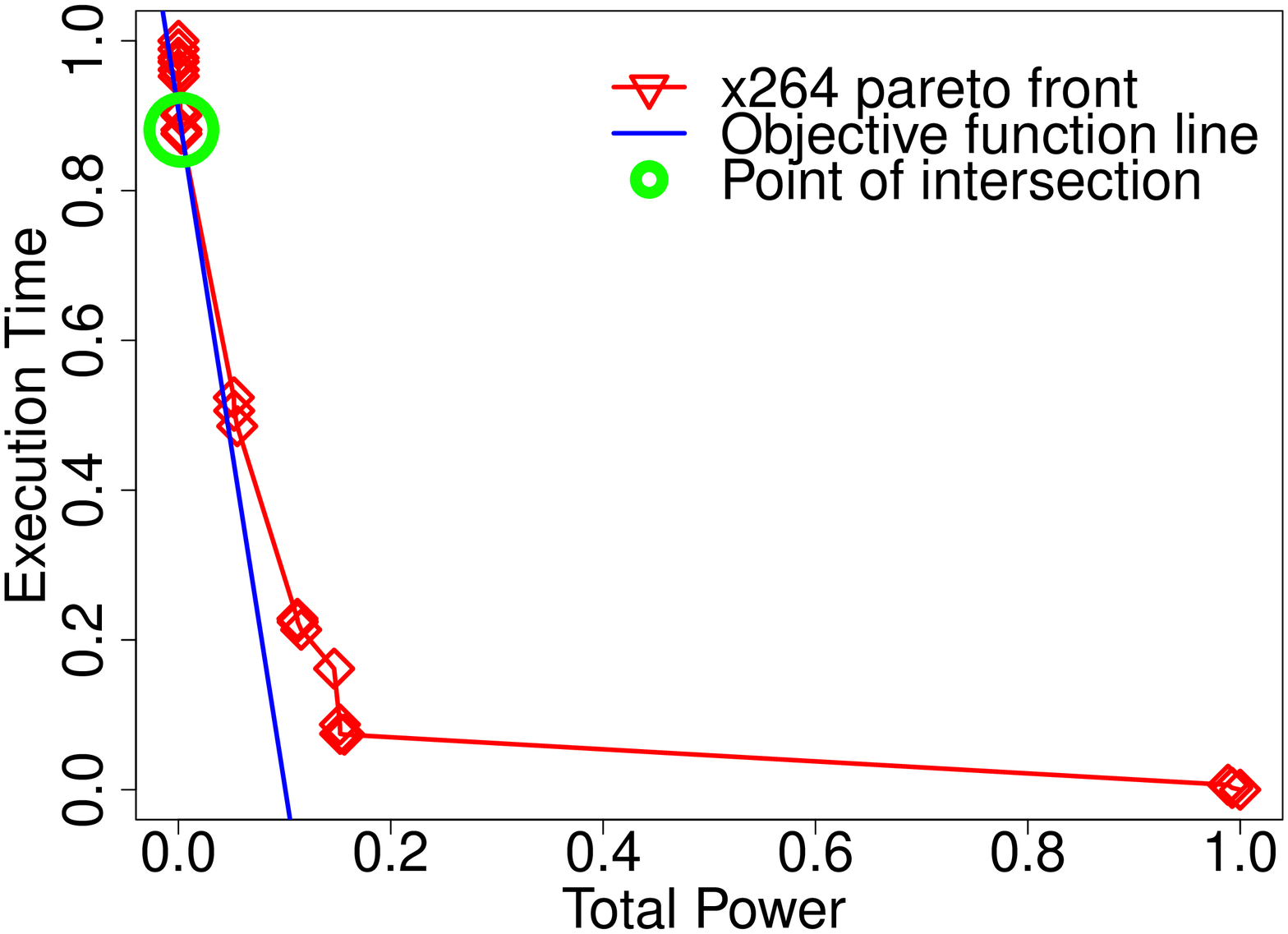}
    \label{figure_smaller_design_space_pareto_tangents_low_power}
    	}
  \hspace{8mm}
  \subfloat[High-performance requirement]
	  {
    	\includegraphics[width = 0.44\textwidth , bb = 0 0 706 536] {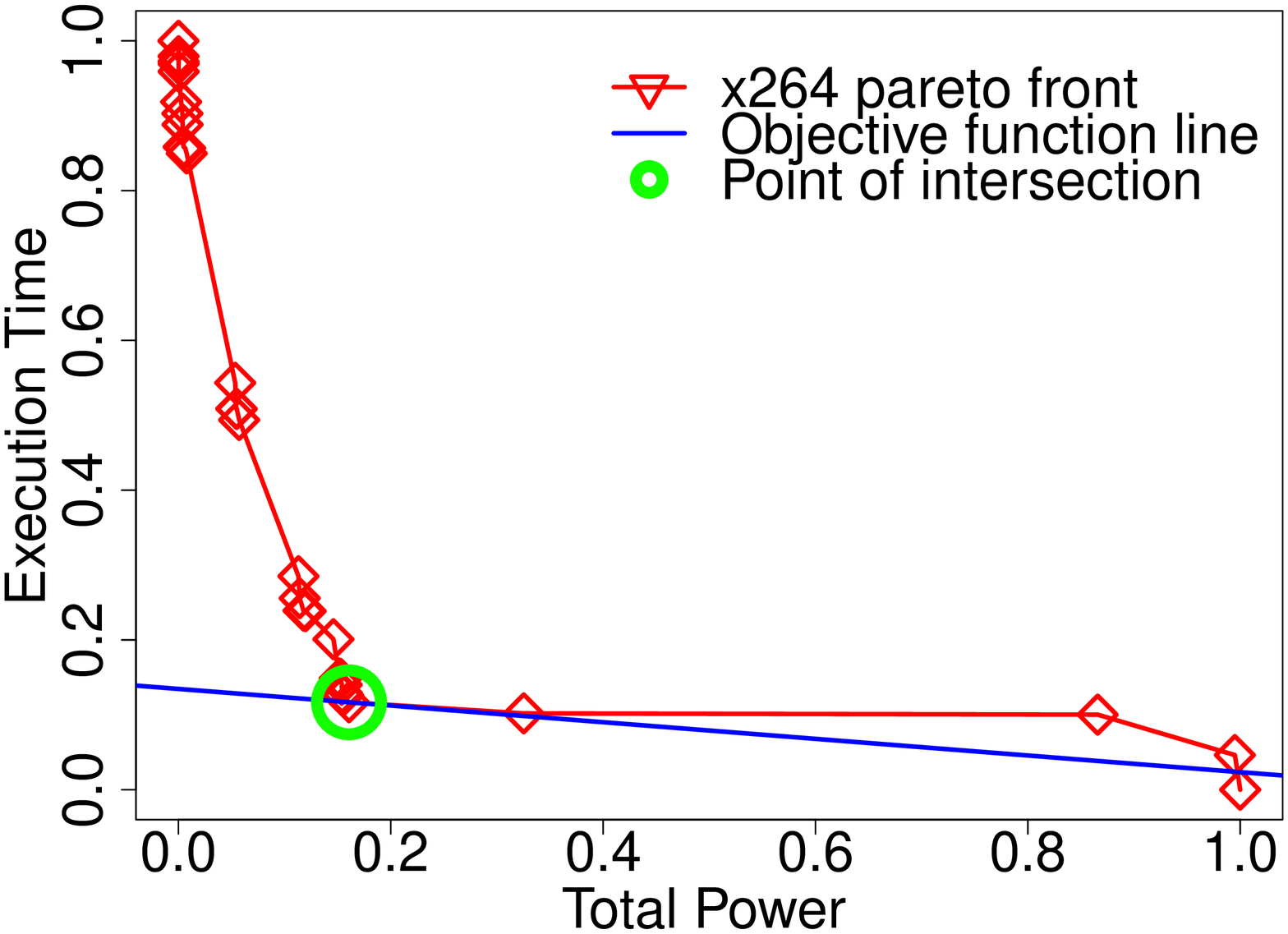}
    \label{figure_smaller_design_space_pareto_tangents_high_performance}
    	}
  \caption{Linear objective function plotted along with the Pareto front for the x264 (PARSEC) benchmark [total power and execution time values are normalized].}
  \vspace{-3mm}
  \label{figure_smaller_design_space_pareto_tangents}
\end{figure*}

Figure \ref{figure_smaller_design_space_pareto_tangents} illustrates how a favorable trade-off solution is selected from the set of Pareto-optimal trade-off points on the Pareto front. The objective function forms a straight line in the power-performance graph with the slope $- w_{\mathcal{P}} / w_{E}$, where $w_{\mathcal{P}}$ and $w_{E}$ are the weights associated with the total power and execution time design metrics, respectively. These weights indicate the preference/weightage of design metrics with respect to each other. This is seen in the objective function line having a smaller intercept on the horizontal axis (Total Power) for low-power requirement and having a smaller intercept on the vertical axis (Execution Time) for high-performance requirement.

Figure \ref{figure_smaller_design_space_pareto_tangents_low_power} reveals that for low-power requirements, the power-performance pair of the form (Total Power (W), Execution Time (ms)), obtained from the point of intersection of the objective line with the Pareto front is (0.904, 68.614). The weight-balanced results obtained from the fully exhaustive search is (0.909, 65.321). This results in a difference of -0.55\% for the total power results and 4.79\% for the execution-time/performance results. The configurations yielding these power-performance pairs for our methodology and the fully exhaustive search both have 2 cores and 1700 MHz frequency. The configurations differ slightly on the cache sizes.

Similarly, from Figure \ref{figure_smaller_design_space_pareto_tangents_high_performance}, the point of intersection of the objective line with the Pareto front for the high-performance requirements case is (1.597, 35.142) and the weight-balanced results obtained from the fully exhaustive search is (1.6, 34.152). This results in a difference of -0.19\% for the total power results and 2.82\% for the execution-time/performance results. Both configurations have 2 cores and 3200 MHz frequency and differ is cache sizes.

Results indicate that our methodology attains results that are consistent with the results obtained from fully exhaustive search results while only exploring 52 configurations (i.e., 1.92\% of the design space) for low-power requirement; and 77 configurations (i.e., 2.85\% of the design space), for high-performance requirement. These results verify that our proposed methodology explores the design space of multicore/manycore processors in a highly efficient manner. Also, the cores count and operating frequency values were determined to be the most significant parameters in the initial (one-shot) phase of our methodology for both low-power and high-performance requirements and hence, were included in the exhaustive search set. Thus, we were able to match these value exactly with the values obtained from the fully exhaustive search.
%
%
\subsubsection{Comparison with Exhaustive Search}
\label{smaller_design_space_exhaustive_search_comparison}
To verify the solution quality (design configuration) obtained by our methodology, we compare the results obtained from our methodology with the results obtained from fully exhaustive search for the given application requirements. Table~\ref{table_results_settings_for_blackscholes} lists details of the design configuration selected by our methodology alongside the one obtained from fully exhaustive search for Blackscholes (PARSEC) test benchmark. The boldface values in the table are parameters settings obtained from fully exhaustive search and the values adjacent to these boldface values are parameter settings obtained from our methodology.
\begin{table}
\caption{Comparison of settings for Blackscholes (PARSEC) Benchmark.}
\label{table_results_settings_for_blackscholes}
\centering
\begin{tabular}[width = \columnwidth]{| l | c | c | c | c |}\cline{2-5}
\multicolumn{1}{c|}{}        &  \multicolumn{2}{c|}{\textbf{High Performance}} & \multicolumn{2}{c|}{\textbf{Low Power}} \\\cline{2-5}
\multicolumn{1}{c|}{}        &  \textbf{EXH} & DSE & \textbf{EXH} & DSE \\\hline
Cores 	                       &  \textbf{8} & 8 & \textbf{2} & 2 \\\hline
Frequency                    &  \textbf{3200} & 3200 & \textbf{1700} & 1700 \\\hline
L1-I Cache Size              &  \textbf{64} & 64 & \textbf{8} & 16 \\\hline
L1-D Cache Size              &  \textbf{32} & 128 & \textbf{16} & 16 \\\hline
L2 Cache Size                &  \textbf{512} & 256 & \textbf{512} & 1024 \\\hline
L3 Cache Size                &  \textbf{4096} & 8192 & \textbf{8192} & 8192 \\\hline
\textbf{Total Power [W]}     &  \textbf{4.309} & 4.549 & \textbf{0.658} & 0.660 \\\hline
\textbf{Execution Time [ms]} &  \textbf{28.153} & 28.139 & \textbf{144.625} & 144.809 \\\hline
\multicolumn{5}{l}{} \\
\multicolumn{5}{l}{\textbf{EXH} - Exhaustive search,} \\
\multicolumn{5}{l}{DSE - Proposed design space exploration methodology}

\end{tabular}
\end{table}
Table \ref{table_results_settings_for_blackscholes} also presents the power-performance values for configurations obtained from fully exhaustive search as well as from our methodology. We observe that for high-performance requirement, our methodology provides a trade-off solution with execution-time/performance within -0.05\% and total power within 5.27\% of the solution obtained from fully exhaustive search. We also note that the configuration obtained for high-performance requirement has a higher core count (i.e., 8 cores) and a higher operating frequency (i.e., 3200 MHz) as compared to the configuration obtained for low-power requirement (i.e., core count of 2 and the operating frequency of 1700 MHz). For low-power requirement, our methodology provides a trade-off solution with total power within 0.3\% and execution time/performance within 0.12\% of the solution obtained from fully exhaustive search. These configuration settings can be explained intuitively. Having a large number of cores working at high frequencies favours high performance whereas having fewer cores operating at low frequencies result in reduced power consumption.

On average, our methodology attains power values within 1.35\% for low-power requirement and the performance values within 3.69\% for high-performance requirement as compared to fully exhaustive search. Our methodology explores 593 configurations in total for all PARSEC test benchmarks (Section~\ref{experimental_setup}) for low-power requirement, and 648 configurations in total for high-performance requirement. The average percentage of the design space explored by our methodology is within 2.74\%--3\%. Hence, our methodology provides a speedup of 35.32$\times$ as compared to fully exhaustive search exploration of the design space. These evaluation results verify that our methodology explores the design space in a highly efficient manner.
%
%
\subsubsection{Comparison with PLATUNE design space exploration methodology}
\label{smaller_design_space_platune_comparison}
The PLATUNE design space exploration methodology is an efficient exploration technique useful for discovering Pareto-optimal configurations. The worst-case running time for this methodology is bounded by $\mathcal{O}(K \times M^{\frac{N}{K}})$, where $K$ denotes the number of strongly connected components in clusters, $N$ denotes the number of parameters and $M$ denotes the upper bound on the number of possible settings for each parameter \cite{tgPLATUNE02}. In the best case, when $K = N$, the running time of their methodology is linear, $\mathcal{O}(N)$. In comparison, the running time of our methodology is always linear, $\mathcal{O}(mT)$ and strictly depends on the exhaustive search limiting factor provided by the system designer. Furthermore, comparing to results reported for PLATUNE in \cite{tgPLATUNE02} and \cite{dsDSEUDOE11}, we observe that our methodology is accurate to within 4\% of results fully exhaustive search exploring less than 3\% of the design space whereas their methodology is accurate to within 8\% of gate-level simulation exploring around 0.2\% of the design space.
%
%
\begin{figure*}[t]
  \centering
  \subfloat[OceanCP]
	  {
	  \includegraphics[width = 0.44\textwidth , bb = 0 0 706 536] {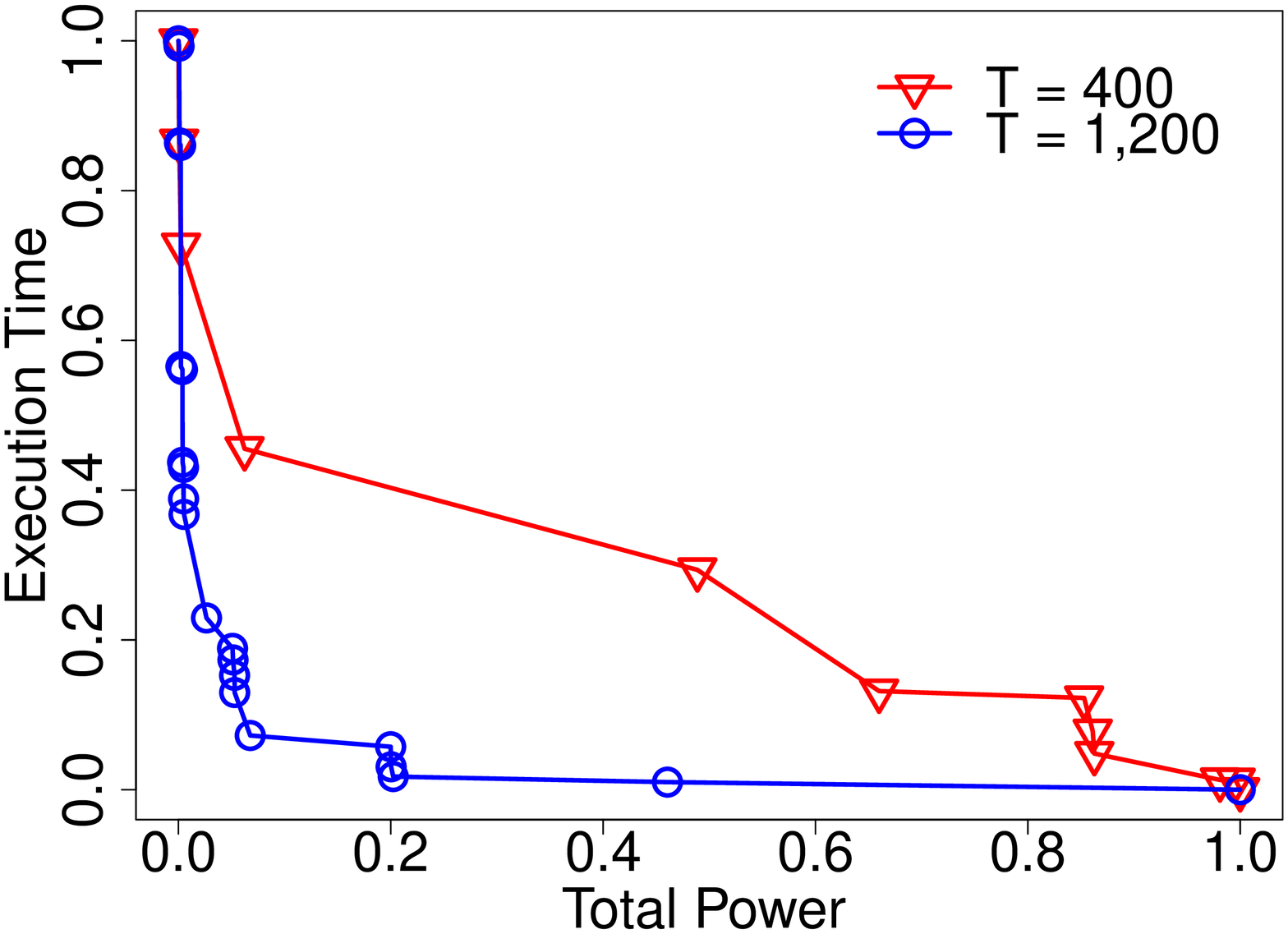}
    \label{figure_splash2_pareto_comparison_ocean_cp}
	  }
  \hspace{8mm}
  \subfloat[Radix]
	  {
	  \includegraphics[width = 0.44\textwidth , bb = 0 0 706 536] {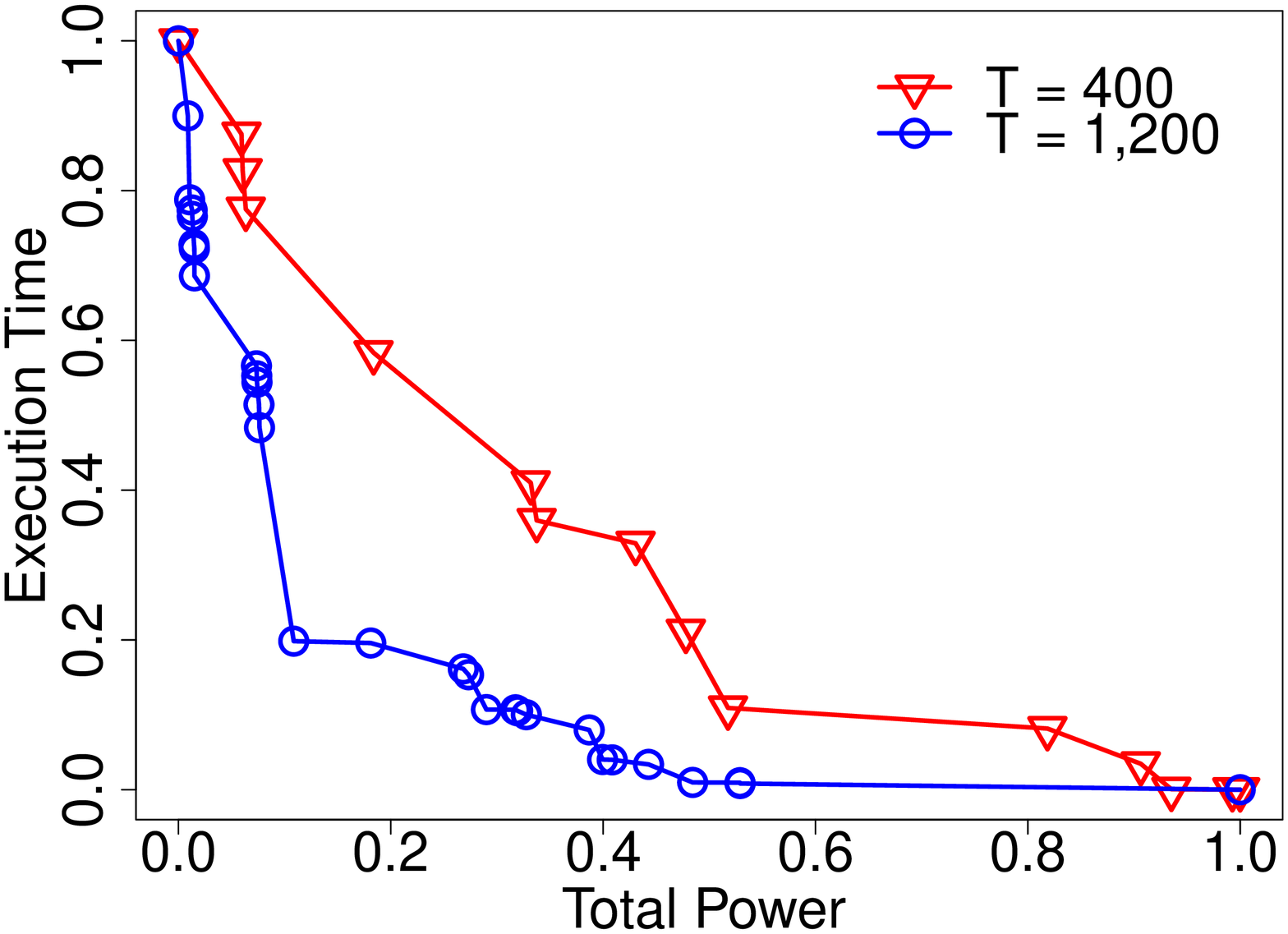}
    \label{figure_splash2_pareto_comparison_radix}
  	  }
  \caption{Comparison of Pareto fronts for exhaustive search threshold factors T = 400 and T = 1,200 for SPLASH-2 test benchmarks [total power and execution time values normalized].}
  \label{figure_splash2_pareto_comparison}
\end{figure*}

\subsection{Evaluation Results for Larger Design Space}
\label{results_larger_design_space}
For the larger search space, we evaluate our methodology by comparing the results (design configurations) obtained by varying  the exhaustive search threshold between two values: $T$ = 400 and $T$ = 1,200. Here, we present a side-by-side comparison of the power-performance values and design configurations (tunable parameter settings) obtained from these two exhaustive search threshold values.
%
%
\subsubsection{Parameter Significance}
\label{larger_design_space_parameter_significance}
\begin{figure}[!t]
    \centering
	\includegraphics[width = 0.44\textwidth , bb = 0 0 706 536] {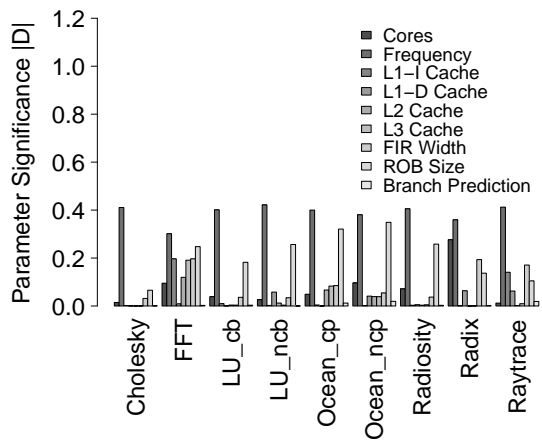}
  	\caption{Significance of tunable parameters for high-performance requirement for SPLASH-2 benchmarks on the larger design space.}
  \label{figure_larger_design_space_parameter_significance}
\end{figure}

Figure \ref{figure_larger_design_space_parameter_significance} shows the normalized values of parameter significance for the SPLASH-2 test benchmarks for high performance requirements. As per intuition, we find that the operating frequency is the most significant parameter out of all the tunable parameters considered performance improvement is the desired design goal. This is consistent with the results obtained for the smaller design space. The second most significant parameter in case of the smaller design space is the cores count, but, with the addition of new tunable parameters - fetch/issue/retire width, reorder buffer size and branch prediction, we find that this is no longer the case. Although, cores count does remain a significant parameter, reorder buffer size takes its place as the second most significant parameter after operating frequency. This is expected because the size of the reorder buffer directly relates to performance, as a higher reorder buffer size allows more instructions to be executed out of order which increases instruction level parallelism \cite{jcROBHPP12} and hence increases performance. For most benchmarks, we observe that the fetch/issue/retire width is the third most significant parameter. We note that a higher fetch/issue/retire width also increases instruction level parallelism which results in performance improvement \cite{mePPIWCMP03}.

%
%
\subsubsection{Pareto Fronts}
\label{larger_design_space_pareto_fronts}
Figure \ref{figure_splash2_pareto_comparison} shows the Pareto fronts for the different exhaustive search threshold factors plotted one over the other for the test benchmarks Ocean\_cp and Radix. From these Pareto fronts, we observe that for higher value of the exhaustive search threshold, the approximation of the Pareto front is better as compared to lower value of exhaustive search threshold. We note that although the Pareto fronts are significantly different, the design configurations (tunable parameter settings) obtained from our methodology for the different threshold factors are fairly consistent.

\begin{table}
\caption{Comparison of settings for SPLASH-2 Benchmark for T = 400 and T = 1,200 for High-performance Requirement}
\label{table_splash2_settings_comparison}
\centering
\begin{tabular}[width = \columnwidth]{| l | c | c | c | c |}\cline{2-5}
\multicolumn{1}{c|}{}         & \multicolumn{2}{c|}{\textbf{Ocean\_cp}} & \multicolumn{2}{c|}{\textbf{Radix}} \\\cline{2-5}
\multicolumn{1}{c|}{}         & \textbf{T = 400} & T = 1,200 & \textbf{T = 400} & T = 1,200 \\\hline
Cores                         & \textbf{2} & 2 & \textbf{4} & 4 \\\hline
Frequency                     & \textbf{3200} & 3200 & \textbf{3200} & 3200 \\\hline
L1-I Cache Size               & \textbf{8} & 128& \textbf{8} & 128 \\\hline
L1-D Cache Size               & \textbf{32} & 128& \textbf{64} & 16 \\\hline
L2 Cache Size                 & \textbf{1024} & 512& \textbf{1024} & 512 \\\hline
L3 Cache Size                 & \textbf{4096} & 8192& \textbf{8192} & 8192 \\\hline
FIR Width                     & \textbf{16} & 4 & \textbf{2} & 2 \\\hline
ROB Size                      & \textbf{256} & 256 & \textbf{64} & 64 \\\hline
Branch Prediction             & \textbf{x2} & x2 & \textbf{x2} & x \\\hline
\textbf{Total Power [W]}      & \textbf{2.622} & 2.312 & \textbf{2.137} & 1.798 \\\hline
\textbf{Execution Time [ms]}  &  \textbf{74.126} & 66.096 & \textbf{7.264} & 7.245 \\\hline
\end{tabular}
\end{table}

Table \ref{table_splash2_settings_comparison} presents a side by side comparison of the configurations obtained from our methodology for different values of exhaustive search threshold, $T$. The boldface values in the tables are the parameter settings obtained from our methodology with $T = 400$ and the values adjacent to these boldface values are the settings obtained from our methodology with $T = 1,200$. We observe that the values obtained using these different exhaustive search threshold factors are highly consistent. The significant parameters - operating frequency, reorder buffer size, fetch/issue/retire width and cores count are mostly similar in the results (design configurations) obtained using these two exhaustive search threshold factors.

The power-performance values for the results (design configurations) obtained from our methodology for the different threshold factors are also presented in Table \ref{table_splash2_settings_comparison}. Before comparing these values, we note that the maximum and average execution times (maximum, average) in ms obtained for all configurations simulated for Ocean\_cp benchmark are (305.84, 149.47) and the same for Radix benchmark are (46.62, 21.25). The execution time values obtained for results (design configuration) obtained for $T = 400$ for the Ocean\_cp benchmark and Radix benchmarks are 74.126~ms and 7.264~ms respectively. The execution time values obtained for the same benchmarks with $T = 1,200$ are 66.096~ms and 7.254~ms respectively. Comparing these values to the maximum and average execution times obtained for all design configurations simulated for these benchmarks, we observe that our methodology determines design configurations with significantly smaller values of execution time (and hence higher performance).

We also note that there is much improvement in the results when using a higher threshold factor. From Table \ref{table_splash2_settings_comparison} we see that increasing the threshold factor from 400 configurations to 1,200 configurations yields an improvement in power and performance by -73.18\% and -10.83\%, respectively, for the Ocean\_cp benchmark and by -15.86\% and -0.26\%, respectively, for the Radix benchmark. This improvement can be explained by the Pareto fronts presented in Figure \ref{figure_splash2_pareto_comparison}. This is also evident from Figure \ref{figure_splash2_pareto_comparison} shows that the approximation of the Pareto front is better when a higher value is used as threshold for exhaustive search. This conforms with the observation presented by Silvano et~al. \cite{csMC10}.

The number of design configurations explored to obtain the results presented in Table \ref{table_splash2_settings_comparison} are 122 and 154 for Ocean\_cp and Radix benchmarks, respectively, for threshold $T = 400$. The same for threshold $T = 1,200$ are 598 configurations for Ocean\_cp and 662 for Radix benchmark. This translates to 0.15\% of the design space for threshold factor $T = 400$ and 0.72\% for threshold factor $T = 1,200$. From these results, it is evident that our methodology determines design configurations which yield significantly better values for design metrics by exploring a very small portion of the design space. Thus, our methodology is a reliable and efficient alternative for design space exploration.
%
%
\vspace{-4mm}
\section{Conclusion and Future Work}
\label{conclusion_and_future_work}
In this paper, we proposed a four-step methodology for efficient design space exploration and tunable parameters optimization for multicore/manycore architectures. The first phase determined good initial settings (within 51.26\% of the best setting) for each of the tunable parameters before beginning the search process. The second phase consisted of a set partitioning algorithm which separated the parameters into different ordered sets based on the significance of the parameters with respect to the objective function and the exhaustive search threshold factor supplied by the system designer. The third and fourth steps consisted of exploring the subsets obtained from the second phase by exhaustive search and greedy search, respectively.

To verify our methodology, we tested it on both small and large processor design spaces. For the smaller design space, containing six tunable parameters, we compared the results (design configurations) obtained from our methodology with the results obtained from fully exhaustive search. The results reveal that our methodology provided results with quality within 1.35\% to 3.69\% of the result quality obtained from fully exhaustive search, while only exploring 2.74\% - 3\% of the design space on average, resulting in a speedup of 35.32$\times$ as compared to fully exhaustive search. For the larger design space, containing nine tunable parameters, we compared the results (design configurations) obtained by varying the exhaustive search threshold between two values, $T = 400$ and $T = 1,200$. We observed that the design configurations obtained for both these exhaustive threshold factors were highly consistent and were significantly better than the average of all other design configurations simulated. Our evaluation also revealed that the approximation of the Pareto front is better for $T = 1,200$ as compared to the same for $T = 400$. This shows that including more number of tunable parameters in the exhaustive search phase of our methodology greatly improves the solution quality, which is consistent with findings of Silvano et~al. \cite{csMC10}. Also, our methodology determined results for the specified application requirements by exploring only 0.15\% - 0.72\% of the larger processor design space. These results verified that out methodology explores the design space efficiently and provides high solution quality.

In the future, we hope to improve on our algorithm by developing a better method for initial one-shot parameter optimization. We intend to investigate full-factorial design method to improve on this step. We plan to improve the set partitioning algorithm by using a better cut-off value for set partition instead of using the exhaustive search threshold factor that we are currently using. Another future goal is to perform comparisons of our methodology against other forms of parameter optimizations using genetic-evolutionary algorithms and machine-learning algorithms.
\vspace{-4mm}
\fussy
}
%
%
{
\balance
\bibliographystyle{IEEEtran}\\
\bibliography{IEEEabrv,TPDS_Final}

%

%
%
\begin{IEEEbiography}[{\includegraphics[width = 1in, bb = 0 -1 105 134]{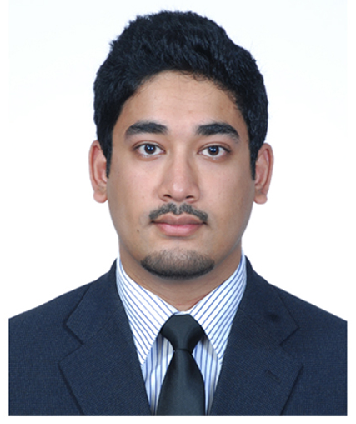}}]{Prasanna Kansakar} is a PhD student in the Department of Computer Science (CS) at Kansas State University (K-State). His research interests include embedded and cyber-physical systems, computer architecture, multicore, and computer security. Kansakar has an MS degree in computer science and engineering from the University of Nevada, Reno (UNR).
\end{IEEEbiography}

\begin{IEEEbiography}[{\includegraphics[width = 1in, bb = 0 -1 722 1010]{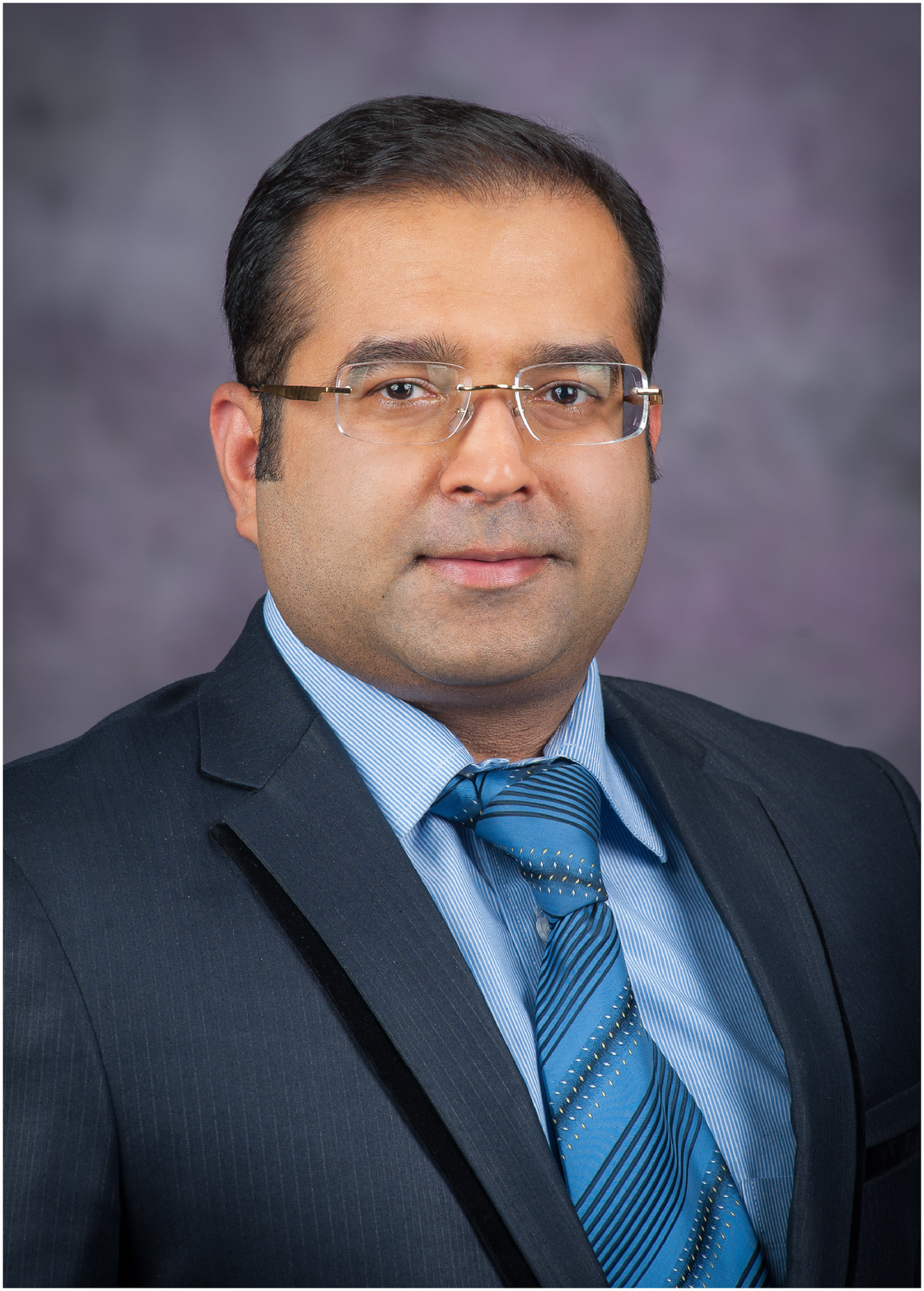}}]{Arslan Munir} is currently an Assistant Professor in the Department of Computer Science (CS) at Kansas State University (K-State). Before joining K-State, he was an Assistant Professor in the Department of Computer Science and Engineering (CSE) at the University of Nevada, Reno (UNR) from July 2014 to June 2017. He was a postdoctoral research associate in the Electrical and Computer Engineering (ECE) department at Rice University, Houston, Texas, USA from May 2012 to June 2014. He received his M.A.Sc. in ECE from the University of British Columbia (UBC), Vancouver, Canada, in 2007 and his Ph.D. in ECE from the University of Florida (UF), Gainesville, Florida, USA, in 2012. He also worked as a visiting graduate research student at the University of Toronto, Toronto, Ontario, Canada for one semester during his Ph.D. From 2007 to 2008, he worked as a software development engineer at Mentor Graphics in the Embedded Systems Division. He was the recipient of many academic awards including the Gold Medals for the best performance in Electrical Engineering, academic Roll of Honor, and doctoral fellowship from Natural Sciences and Engineering Research Council of Canada (NSERC). He received a best paper award at the IARIA International Conference on Mobile Ubiquitous Computing, Systems, Services and Technologies in 2010. One of his research paper was also selected as the Best Paper Finalist (one of the top three papers) at the IEEE International Conference on Application-specific Systems, Architectures and Processors (ASAP) in 2017. His current research interests include embedded and cyber-physical systems, secure and trustworthy systems, computer architecture, multicore, parallel computing, distributed computing, and fault tolerance.
\end{IEEEbiography}

}
\end{document}